\documentclass[]{pasj00}

\newcommand{\ergcms}{{\rm ergs}\ {\rm cm}^{-2}\ {\rm s}^{-1}}

\Received{$\langle$reception date$\rangle$}
\Accepted{$\langle$acception date$\rangle$}
\Published{$\langle$publication date$\rangle$}
\SetRunningHead{T. Yuasa et al.}{Suzaku Detection of Extended/Diffuse Hard X-Ray Emission from the Galactic Center}

\usepackage{times}

\begin{document}

\title{Suzaku Detection of Extended/Diffuse Hard X-Ray Emission\\ from the Galactic Center}
\author{
Takayuki \textsc{Yuasa},\altaffilmark{1}
Ken-ichi \textsc{Tamura},\altaffilmark{2,1}
Kazuhiro \textsc{Nakazawa},\altaffilmark{1}
Motohide \textsc{Kokubun},\altaffilmark{2}
Kazuo \textsc{Makishima},\altaffilmark{1,3}\\
Aya \textsc{Bamba},\altaffilmark{2}
Yoshitomo \textsc{Maeda},\altaffilmark{2}
Tadayuki \textsc{Takahashi},\altaffilmark{2,1}
Ken \textsc{Ebisawa},\altaffilmark{2,1}
Atsushi \textsc{Senda},\altaffilmark{3}\\
Yoshiaki \textsc{Hyodo},\altaffilmark{4}
Takeshi Go \textsc{Tsuru},\altaffilmark{4}
Katsuji \textsc{Koyama},\altaffilmark{4}
Shigeo \textsc{Yamauchi},\altaffilmark{5}
Hiromitsu \textsc{Takahashi}\altaffilmark{6}
}
\altaffiltext{1}{Department of Physics, School of Science, University of Tokyo, 7-3-1 Hongo, Bunkyo-ku, Tokyo 113-0033}
\altaffiltext{2}{Department of High Energy Astrophysics, Institute of Space and Astronautical Science (ISAS), \\Japan Aerospace Exploration Agnency (JAXA), 3-1-1 Yoshinodai, Sagamihara, Kanagawa 229-8510}
\altaffiltext{3}{Cosmic Radiation Laboratory, The Institute of Physical and Chemical Research (RIKEN), 2-1 Hirosawa, Wako, Saitama 351-0198}
\altaffiltext{4}{Department of Physics, Graduate School of Science, Kyoto University, Kita Shirakawa Oiwake-cho, Sakyo-ku, Kyoto 606-8502}
\altaffiltext{5}{Faculty of Humanities and Social Sciences, Iwate University, 3-18-34 Ueda, Morioka, Iwate 020-8550}
\altaffiltext{6}{Department of Physical Science, Hiroshima University, 1-3-1 Kagamiyama, Higashi-Hiroshima, Hiroshima 739-8526}
\email{yuasa@amalthea.phys.s.u-tokyo.ac.jp}

\maketitle

\begin{abstract}
Five on-plane regions within $\pm \timeform{0\circ.8}$ of the Galactic center
were observed with the Hard X-ray Detector (HXD) and the X-ray Imaging Spectrometer (XIS)  onboard Suzaku.
From all  regions,
significant hard X-ray emission was detected  with HXD-PIN up to 40 keV,
in addition to the extended plasma emission which is dominant in the XIS band.
The hard X-ray signals are inferred to come
primarily from a spatially extended source,
rather than from a small number of  bright discrete objects.
Contributions to the HXD data from catalogued X-ray sources,
typically brighter  than 1 mCrab,
were estimated and removed using information from Suzaku and other satellites.
Even after this removal,
the hard X-ray signals remained significant,
exhibiting a typical 12--40 keV surface brightness of 
$4\times10^{-10}~\ergcms~\mathrm{deg}^{-2}$
and power-law-like spectra with a photon index of 1.8.
Combined fittings to the XIS and HXD-PIN spectra confirm
that a separate hard tail component is
superposed onto the hot thermal emission,
confirming a previous report based on the XIS data.
Over the 5--40 keV band,
the hard tail is spectrally approximated
by a power law of photon index $\sim 2$,
but better by those with somewhat convex shapes.
Possible origins of the extended hard X-ray emission are discussed.
\end{abstract}

%%%%%%%%%%%%%%%%%%%%%%%%%%%%%%%%%%%%%%%%%%%%
%Introduction
%%%%%%%%%%%%%%%%%%%%%%%%%%%%%%%%%%%%%%%%%%%%
\section{Introduction}
Extended X-ray emission associated with our Galaxy
has been observed for more than 20 years
(\cite{wor82}; \cite{iwa82}; \cite{war85}; \cite{koy86}), 
and has been revealed to consist of three distinct spatial components;
the Galactic ridge component (\cite{koy86}; \cite{kan97}), 
the Galactic bulge components (\cite{yam93}; \cite{kok01}),
and  the Galactic center (GC) component
discovered with Ginga (\cite{koy89}; \cite{yam90}).
Over a typical energy range of  $2-10$ keV,
all the three emission components are
spectrally dominated by thermal emission from hot plasmas
with a temperature of $\sim 10^8$ K,
as evidenced by strong K-shell emission lines from highly ionized iron in the spectra
\citep{koy86,yam93,kan97,kok01}.

The overall phenomenon has been interpreted either as truly diffuse 
emission permeating the interstellar space (\cite{sug01}; \cite{ebi01}), 
or as superposition of a large number of 
unresolved discrete (mostly point-like) X-ray sources,
such as cataclysmic variables (CVs) and RS CVn type binaries \citep{rev06}.
However, either interpretation has problems. 
If the emission is from truly diffuse plasma, 
the measured temperature ($\sim5-10$ keV) 
is inferred to significantly exceed the gravitational 
escape temperature of the  Galaxy ($\sim0.5$ keV),
and the pressure to significantly exceed 
that of any known interstellar energy component.
If instead the plasma were escaping freely from the disk,
a large input energy would be required 
to supply the plasma and to sustain the emission within the escaping time scale of the plasma of $\sim10^5$ yr.
The other scenario invoking discrete sources also has a serious difficulty, 
in that deep observations with Chandra resolved only $10-30$\% 
of the total emission from the ridge (\cite{ebi01}; \cite{rev07}) and the GC \citep{mun04} 
into point sources:
to account for the remainder,
a new class of much dimmer but more numerous X-ray point sources are needed.

A direct comparison of the emission spectra of the three spatial components with one another, and with those of the candidates of unresolved sources, 
will provide an important clue to the origin of the extended Galactic X-ray emission. 
Actually, the ridge and bulge emissions have been extensively studied 
in energies both below and above 10 keV, 
incorporating imaging (e.g., \cite{kan97}) and 
collimated (e.g., \cite{yam97,kok01}) instruments, respectively. 
As a result, the spectra of the X-ray emission filling these two regions have been confirmed 
to exhibit a clear spectral excess in high energies, or a hard tail, 
above the thermal emission (\cite{yam97}; \cite{val98}; \cite{val00}; \cite{kok01}).
The excess has been taken as evidences for ongoing
particle acceleration in the interstellar space \citep{yam97},
or alternatively, for a significant population of
numerous hard X-ray sources such as CVs \citep{rev06}.

In the GC region, the extended thermal emission has  been studied 
extensively in energies below 10 keV \citep{sid99a,mun04},
and the recent Suzaku X-ray Imaging Spectrometer (XIS) 
observations have revealed a strong spectral 
hard tail to accompany the GC emission as well  \citep{koy07a}.
Nevertheless, its direct confirmation in energies above $\sim 10$ keV
has so far been unavailable,
because studies with collimated or coded-mask instruments
are severely hampered by the high surface density 
of bright X-ray point sources around the GC.

The silicon PIN diodes (hereafter HXD-PIN) of the Hard X-ray Detector (HXD; \cite{tak07}; \cite{kok07}) 
onboard the Suzaku satellite \citep{mit07}
has a tightly collimated field of view (FOV) of $34\prime\times34\prime$ (FWHM)
with the lowest detector background ever achieved,
and enables to measure the hard tail component of 
the extended GC  emission in energies above 10 keV, 
without being hampered by contamination of bright point sources. 
In fact, the solid angle of HXD-PIN is $\sim4.5$ and $\sim2.5$ 
times smaller than that of the BeppoSAX PDS and the RXTE HEXTE, respectively.
In the present paper, we report on the HXD-PIN detection of apparently extended bright 
hard X-ray emission  from the GC region.

%%%%%%%%%%%%%%%%%%%%%%%%%%%%%%%%%%%%%%%%%%%%
%Observation
%%%%%%%%%%%%%%%%%%%%%%%%%%%%%%%%%%%%%%%%%%%%
\section{Observation}\label{sec:obs}
We observed 5 different regions around the GC ($\timeform{-0\circ.5}<l<\timeform{+0\circ.5}$) 
with the XIS and the HXD onboard  Suzaku, 
in 2005 September and October, 
as well as in 2006 February, March and September, 
each for an exposure of 30$-$70 ks. 
As listed in table \ref{tab:obslog}, 
we hereafter call these observed regions A, B, C, D and E. 
Region A and B, the nearest two to the GC, were observed three times and twice respectively, 
so we distinguish those observations by putting subscripts.

In 2005 September, additional three short offset observations were conducted, 
to obtain information on known bright X-ray point sources 
that lie outside the XIS field of view 
but inside the PIN FOV during the relevant observations. 
These offset data allow us to estimate to what extent 
they contaminate the HXD data from regions A through E. 
The point sources targeted in the offset observations are 
1A 1742$-$294, KS 1741$-$293 and 1E 1743.1$-$2843; 
the net exposure of the XIS for these point sources were
 5.2 ks, 4.8 ks and 4.2 ks, respectively (table \ref{tab:obslog}).

\begin{table*}
\caption{Observations of the Galactic Center region.}\label{tab:obslog}
\begin{center}
\begin{tabular}{lccccc}
\hline\hline
Name & Sequence & \multicolumn{2}{c}{Position\footnotemark[$*$]} & Start & Exp.\footnotemark[$\dagger$]\\
	& number	&	$l$ ($^\circ$)	&	$b$ ($^\circ$)	&	(UTC)	& (ks)\\
\hline
Region A$_1$ & 100027010 & 0.06 & 0.07 & 2005-09-23 07:27 & 31.1 \\
Region A$_2$ & 100037040 & 0.06 & 0.07 & 2005-09-30 07:43 & 33.3 \\
Region A$_3$ & 100048010 & 0.06 & 0.07 & 2006-09-08 02:32 & 38.2 \\
Region B$_1$ & 100027020 & -0.25 & 0.05 & 2005-09-24 15:22 & 30.3 \\
Region B$_2$ & 100037010 & -0.25 & 0.05 & 2005-09-29 05:21 & 32.8 \\
Region C & 500005010 & 0.43 & 0.12 & 2006-03-27 23:48 & 71.7 \\
Region D & 100037060 & 0.64 & 0.10 & 2005-10-10 12:28 & 56.6 \\
Region E & 500018010 & -0.57 & 0.09 & 2006-02-20 13:01 & 38.3 \\
\hline \\

\hline\hline
Name & Sequence & \multicolumn{2}{c}{Target Position\footnotemark[$\ddagger$]} & Start & Exp.\footnotemark[$\S$]\\
	& number	&	$l$  ($^\circ$)	&	$b$	 ($^\circ$) &	(UTC)	& (ks)\\
\hline
Offset1$\_$1 (1A 1742$-$294) & 100027030 & -0.441 & -0.389 & 2005-09-24 12:06 & 2.0 \\
Offset1$\_$2 (1A 1742$-$294) & 100037020 & -0.441 & -0.389 & 2005-09-30 04:35 & 3.2 \\
Offset2$\_$1 (KS 1741$-$293) & 100027040 & -0.416 & -0.087 & 2005-09-24 13:44 & 1.9 \\
Offset2$\_$2 (KS 1741$-$293) & 100037030 & -0.416 & -0.087 & 2005-09-30 06:06 & 2.9 \\
Offset3$\_$1 (1E 1743.1$-$2843) & 100027050 & 0.251 & -0.026 & 2005-09-25 18:00 & 1.9 \\
Offset3$\_$2 (1E 1743.1$-$2843) & 100037050 & 0.251 & -0.026 & 2005-10-01 06:57 & 2.3 \\
\hline
\multicolumn{6}{@{}l@{}}{\hbox to 0pt{\parbox{180mm}{\footnotesize
\vspace{0.2cm}
\footnotemark[$*$]FOV center position.\par
\footnotemark[$\dagger$]Effective exposure of the PIN data.\par
\footnotemark[$\ddagger$]The positions of targeted point sources.\par
\footnotemark[$\S$]Effective exposure of the XIS data.
}\hss}}
\end{tabular}
\end{center}
\end{table*}

%%%%%%%%%%%%%%%%%%%%%%%%%%%%%%%%%%%%%%%%%%%%
%Data Reduction
%%%%%%%%%%%%%%%%%%%%%%%%%%%%%%%%%%%%%%%%%%%%
\section{Data Reduction}\label{sec:dr}
The reduction and analysis of the present data were carried out using the HEADAS  software version 6.1.2. The data produced by Suzaku Data Processing version 1.3 were analyzed. In the spectral fitting, we used \texttt{xspec} version 11.3. The utilized PIN response matrix file versions were 2006-08-09 and 2006-08-14 for uniform emission and a point source, respectively. The XIS response matrices and auxiliary response files were calculated using \texttt{xisrmfgen} and \texttt{xissimarfgen} (version 2006-11-26).

We removed the XIS and HXD data obtained while the elevation angle is less than 5$^\circ$ (ELV$<5^\circ$) and the spacecraft is in the South Atlantic Anomaly. Additionally, we imposed the condition that the day earth elevation angle is less than 20$^\circ$ (DYE$\_$ELV$<20^\circ$) for the XIS data. The HXD data acquired in regions of low cutoff rigidity (COR$<8$ GV) were also excluded.

In analyzing the HXD data, we extracted the signals from all 64 HXD-PIN detectors 
in each observation except for Region A$_3$. 
Because of a break-down like event in a silicon PIN detector 
which took place  in 2006 May, the high voltage supplied to 16 PIN detectors
(including that particular one) was changed  from 500 V to 400 V.
The third observation of Region A (Region A$_3$) was 
conducted after this operation.
Because the response under the reduced high voltage 
is yet to be calibrated, we limited the present analysis of the A$_3$ data  
to signals  from the other 48 PIN detectors, 
resulting in 25\% reduction in the effective area. 
In the present study,  we did not use the GSO data 
because of its lower signal to noise ratio than that of PIN by about an order of magnitude.

We estimated non X-ray background (NXB) of the XIS 
using night-earth data sorted by cut off rigidity. 
That of the HXD was estimated by a synthetic model \citep{kok07}, 
which mainly relies on the count rate of the PIN upper discriminator 
(i.e. count rate of cosmic ray particles) 
and utilizes night-earth data as background templates. 

The GC region hosts a large number of discrete X-ray sources.
In the following analysis, however, we did not remove any point sources, 
and extracted events from whole area of the XIS CCDs,
excluding the corners illuminated by calibration radioisotope signals. 

\begin{figure*}
  \begin{center}
    \FigureFile(150mm,50mm){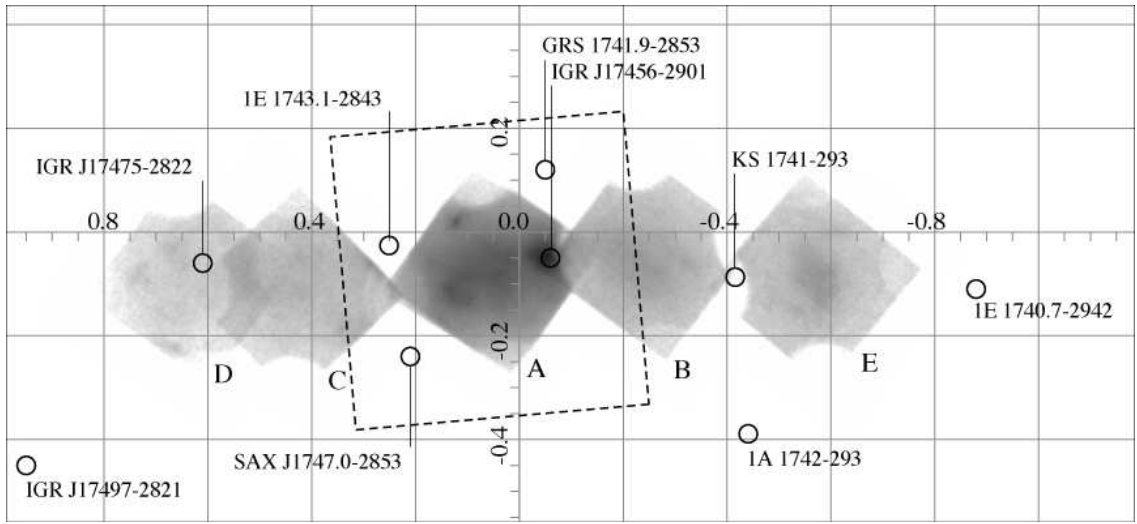}
  \end{center}
  \caption{A $2-10$ keV background-inclusive mosaic image of the Galactic 
  center region taken with the XIS. The image is corrected for 
  neither the exposure nor vignetting. 
  The overlaid dotted square is the FWHM FOV of the PIN detector during 
  the Region A observation. Black circles indicate the catalogued bright point sources 
  that can contaminate the PIN signals (\S\ref{subsec:subtraction}).
  }
  \label{fig:gcimage}
\end{figure*}

Figure \ref{fig:gcimage} shows a mosaic image of the GC region, 
constructed from the five pointings, 
using $2-10$ keV data from all the four XIS CCDs. 
All the imaging area of the CCD chips is used, 
except for the two corners of each camera irradiated by calibration isotopes.

%%%%%%%%%%%%%%%%%%%%%%%%%%%%%%%%%%%%%%%%%%%%
%Analysis and Results
%%%%%%%%%%%%%%%%%%%%%%%%%%%%%%%%%%%%%%%%%%%%
\section{Analysis and Results}

\subsection{Light curves}
\begin{figure}
  \begin{center}
    \FigureFile(80mm,30mm){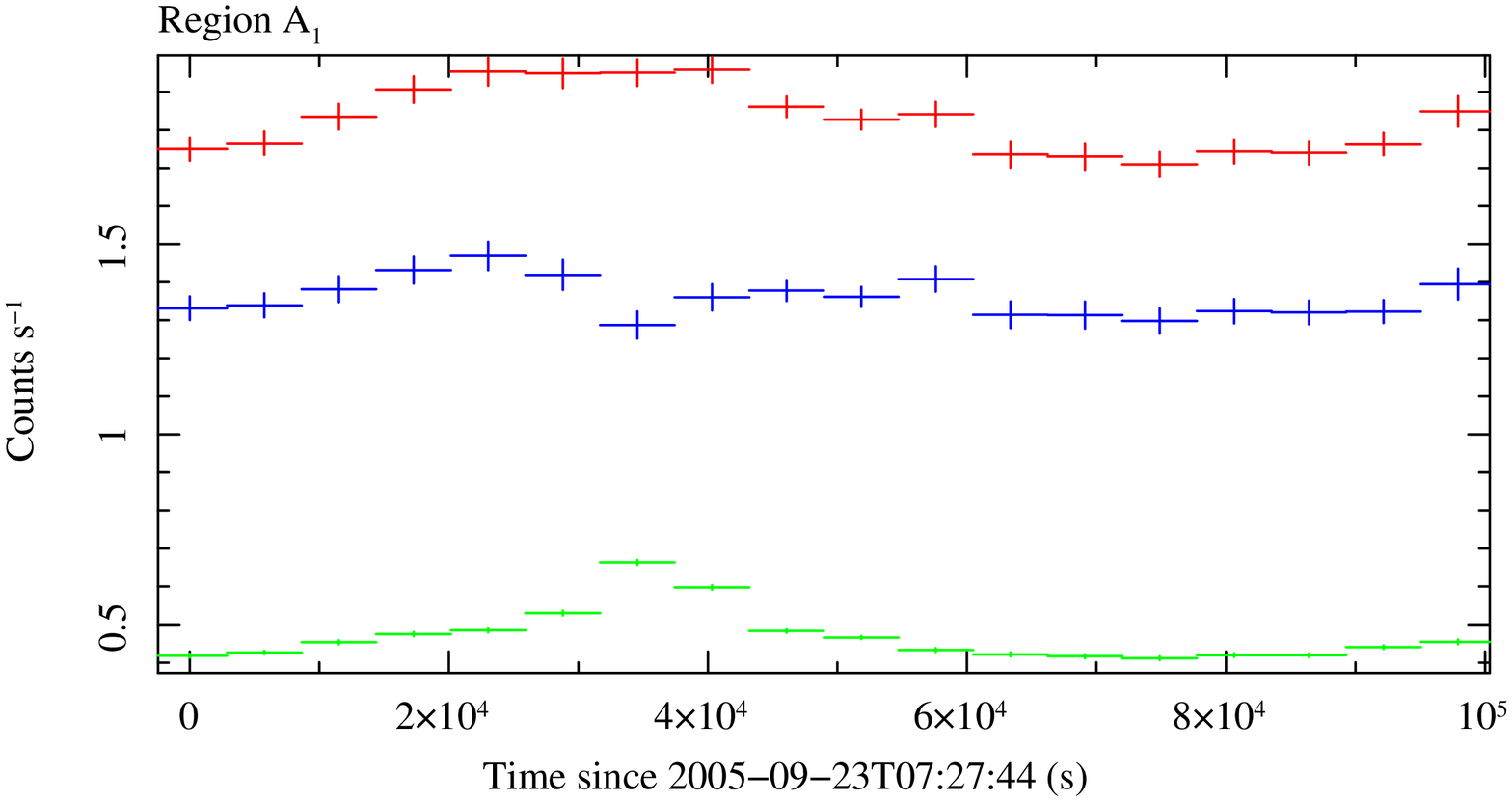}
    \FigureFile(80mm,30mm){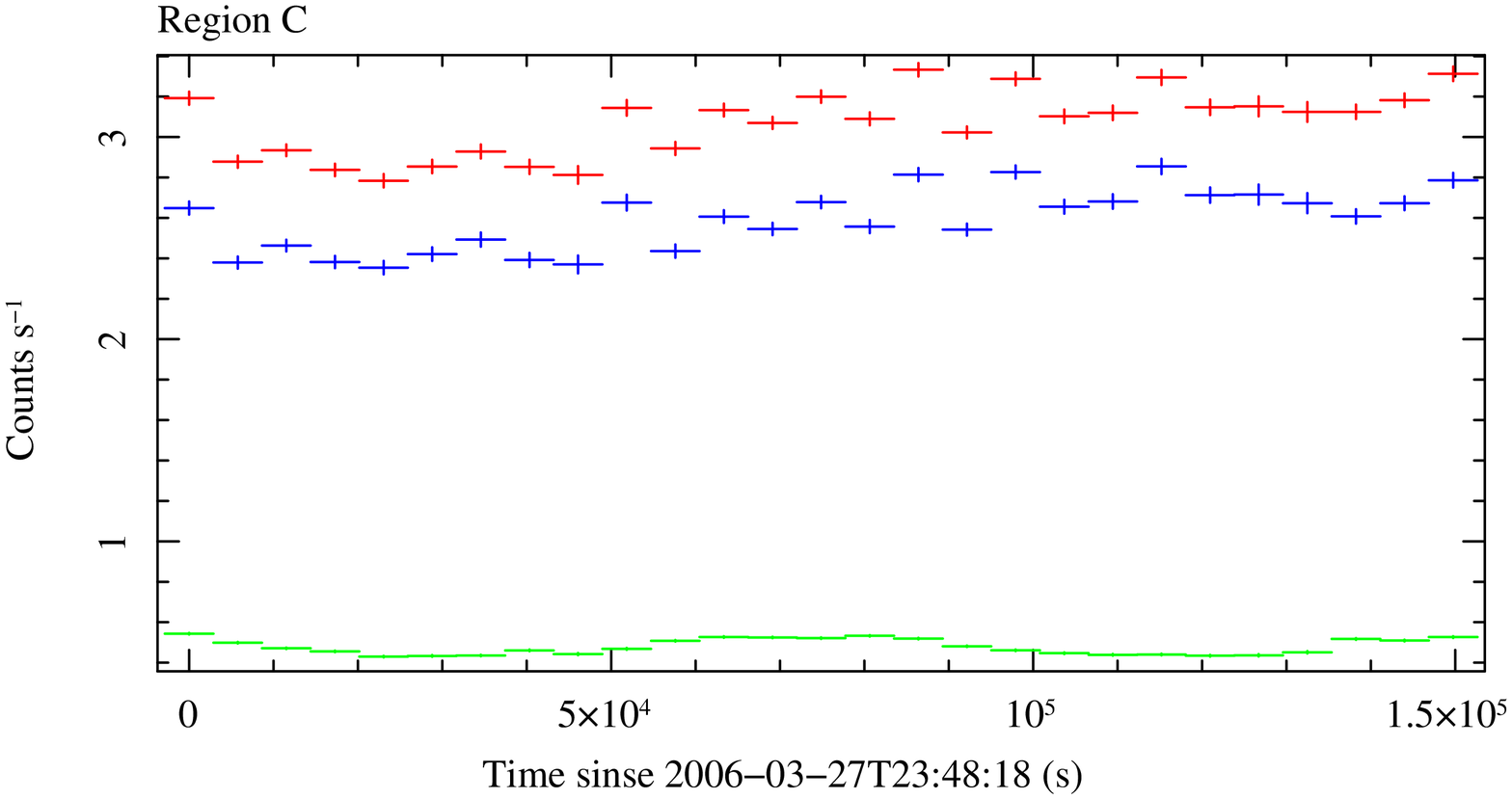}
  \end{center}
  \caption{HXD-PIN light curves of Region A$_1$ and Region C in the 10$-$40 keV band, 
  binned into 5760 s. Each plot shows raw counts (red), the modeled NXB (green), 
  and the NXB-subtracted signals (blue). }\label{fig:lc}
\end{figure}

Figure \ref{fig:lc}  shows light curves of the raw 
$10-40$ keV HXD-PIN count rates of Region A$_1$ and Region C, 
binned into the orbital period of the spacecraft, 5760 s. 
The modeled NXB and the NXB-subtracted count rates are also plotted. 
The raw counts were corrected for dead time fraction in each bin, 
which was calculated from the ``pseudo-event'' rates 
(\cite{tak07}; \cite{kok07}) as about $5-9\%$ . 
The NXB counts (and hence the raw counts too) 
exhibit periodic variations due to activation in the South Atlantic Anomaly (SAA)
and subsequent decays of the produced radioactive isotopes.
Since the reproducibility of the NXB model decrease 
(overestimating the NXB counts) slightly when 
the satellite orbit crosses the SAA,
the NXB-subtracted signal count rate in Region A$_1$ are 
also undulated around $(2-5)\times10^4$~s. However in other
time regions, it is rather constant within statistical errors. 

We likewise confirmed that the NXB-subtracted signal counts of all other regions, 
except for that of Region C, are rather constant during the present observations. 
However, as shown in figure \ref{fig:lc} (bottom), 
the count rate from Region C increased 
by $\sim15\%$ in $\sim1.5\times10^5$ s,
presumably because a point source SAX J171747.0$-$2853, 
located outside the XIS FOV but inside the PIN FOV, 
varied during the present observation (\S\ref{subsection:regionc}).

\subsection{Spectra}\label{subsection:spectra}
\begin{figure}
  \begin{center}
    \FigureFile(80mm,50mm){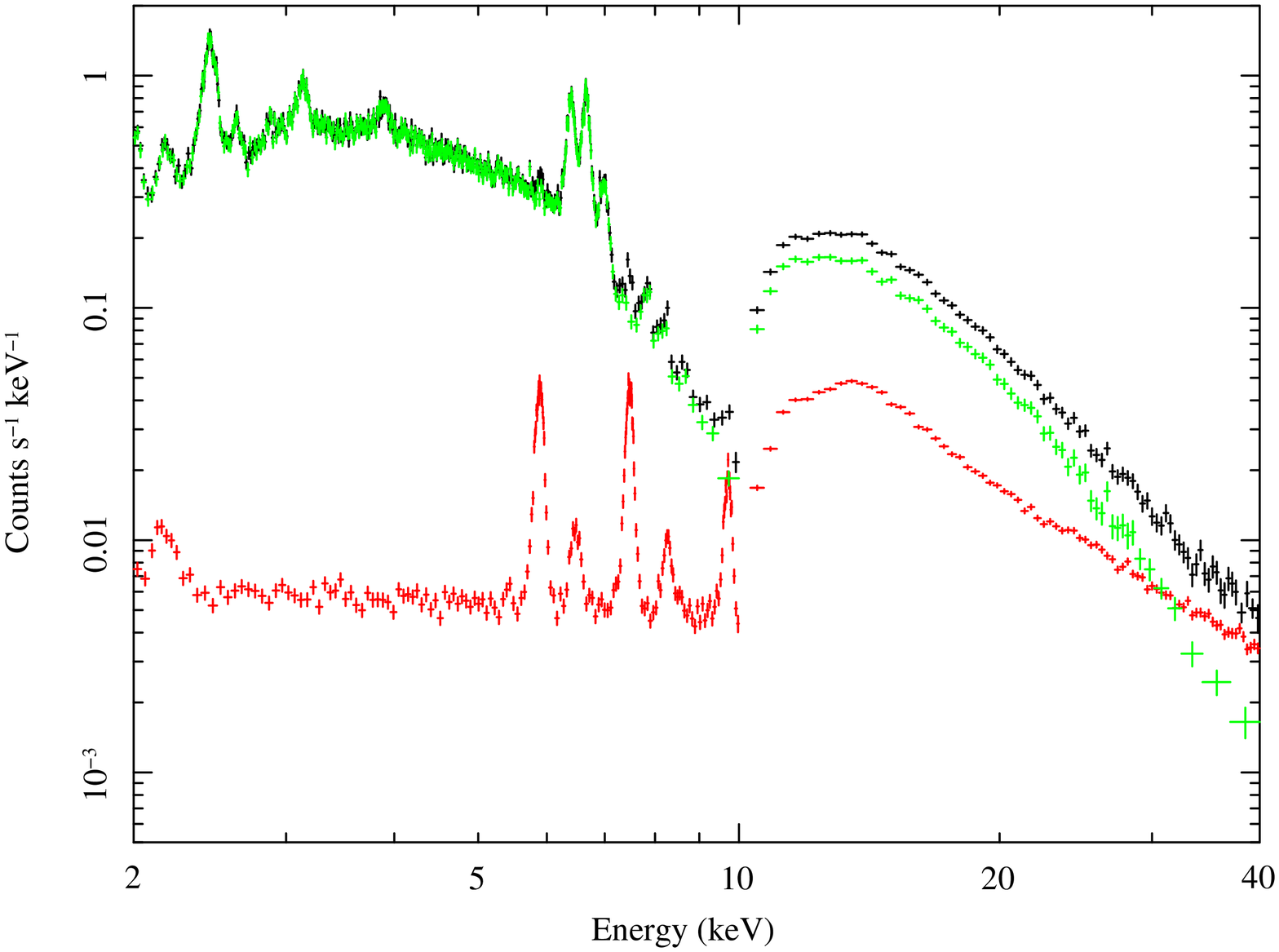}
    \FigureFile(80mm,50mm){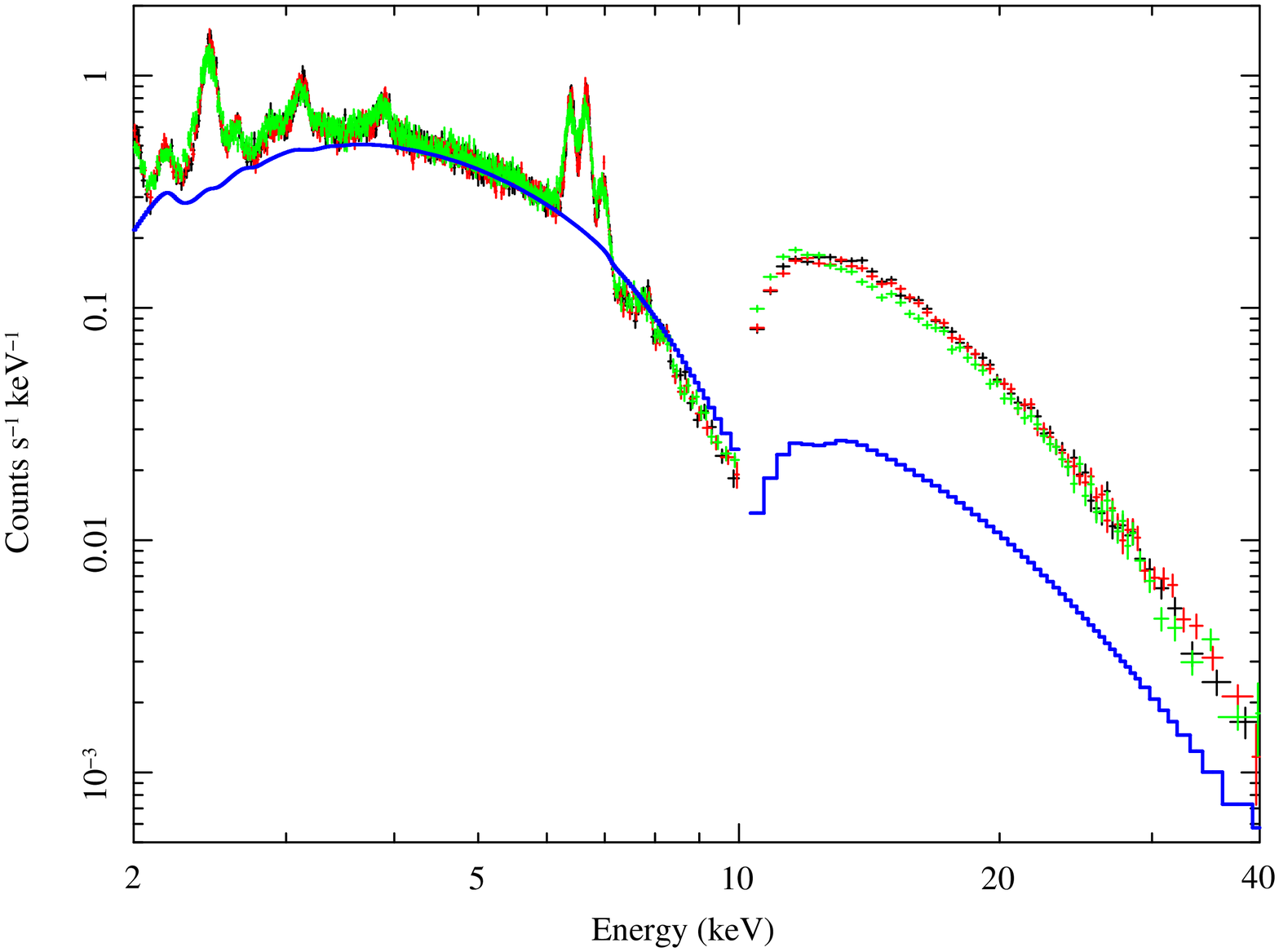}
  \end{center}
  \caption{Top : Wide-band spectra of Region A$_1$, taken with XIS0 and PIN. 
  Raw (black), NXB (red), and NXB-subtracted (green) spectra are plotted. 
  Bottom : The NXB-subtracted spectra of XIS0 and PIN obtained in the first, second, 
  and third  observations of Region A,  shown in black, red, and green crosses, respectively. 
 The PIN spectrum of the third observation is scaled to 4/3 of the original one (see text).
 Blue solid line is a model spectrum of a 5 mCrab point source 
 which has a single power-law spectrum with $\Gamma=2.1$, 
 absorbed by a hydrogen column of $N_\mathrm{H}=6\times10^{22}~\mathrm{cm}^{-2}$.}
  \label{fig:gcsrc1}
\end{figure}

Figure \ref{fig:gcsrc1} shows a typical example of the wide-band GC spectra, 
obtained from Region A$_1$ with XIS0 ($2-10$ keV) and 64 PIN diodes ($10-40$ keV).
We limit the present analysis of the XIS data to energies above 2 keV, 
to avoid the contribution from a softer thermal component \citep{koy07a}.
The NXB subtracted HXD signal is so significant 
that it exceeds the NXB over $\sim10-30$ keV band. 
Because the reproducibility of the PIN NXB is $3-5\%$ \citep{kok07}, 
we can claim the detection of a significant hard X-ray flux, 
at least up to $\sim40~\mathrm{keV}$ in this particular case.
 For detailed analysis of the XIS data of these regions, 
\citet{koy07a} may be referred to.

Region A was observed three times, 
with a time separation of a week between the first two observations, 
while a year till the last one. 
The NXB subtracted spectra of those three observations 
are shown in the bottom panel of figure \ref{fig:gcsrc1}. 
The data from the third pointing are corrected for 
the 25\% loss of the PIN effective area (\S3). 
The three spectra turned out to be very similar, 
with the $10-40$ keV PIN count rate (after the background subtraction) 
in agreement within $\sim10\%$.
Therefore, across the three observations,
the hard X-ray emission from this region 
exhibited only minor variations, if any.

In the same panel, a model spectrum predicted for a 5 mCrab point source 
(a single power law of photon index $\Gamma=2.1$)
located at the FOV center is also plotted. 
The observed PIN signals much exceed 
what is expected when a single point source, 
which roughly matches the XIS counts, 
was present at the FOV center of the XIS and PIN. 
We can explain this flux discrepancy in two alternative ways. 
One is to assume that the emission is extended, 
and hence HXD-PIN, 
which has a wider FOV than the XIS, received higher counts than the XIS. 
The other is to consider that bright point sources, 
which reside outside the XIS FOV but inside the PIN FOV, 
were contributing to the PIN flux.

\begin{figure}
  \begin{center}
    \FigureFile(85mm,50mm){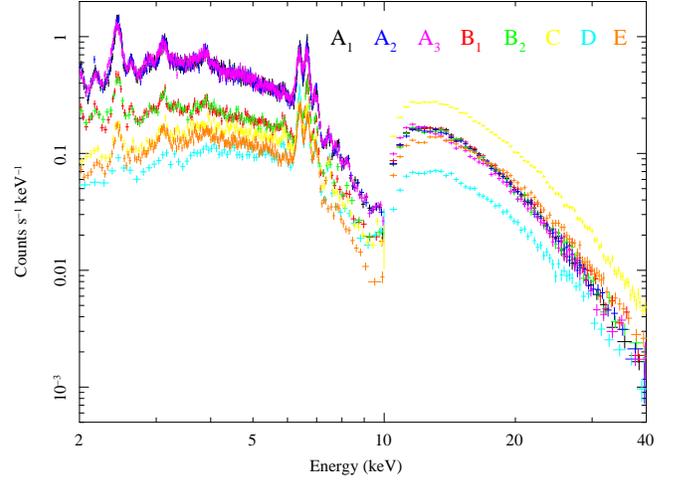}
  \end{center}
  \caption{Wide-band spectra of the 5 regions.
  Data obtained with XIS0 ($2-10$ keV) and HXD-PIN ($10-40$ keV) are plotted
  in the same manner as figure~\ref{fig:gcsrc1}.
  Different observations are indicated by different colors,
  as specified in the figure.
 }
  \label{fig:gc5regions}
\end{figure}

In order to evaluate the above two alternatives, 
we present in figure \ref{fig:gc5regions} 
the background-subtracted wide-band spectra from all regions. 
We fitted these PIN spectra with a single power-law model.
Although the emission may well be extended, 
we employed the response file for a point source located 
at the PIN optical axis. This is to make it easier to compare the detected flux with
that of known bright point sources, which can contaminate the PIN signals
(see below).
The derived best fit parameters are listed in table \ref{tab:pin_powerlaw_hxdnom}. 
Among the 5 regions, except in Region C and D, 
the PIN signals have very similar intensities,
as well as comparable spectral slopes.
Even though there should be some point sources 
outside the XIS FOV but inside the HXD FOV,
it would be rather difficult for them to be arranged 
in such a way that they contribute almost equally to Region A, B and E. 
Furthermore, in the $10-20$ keV band, the PIN signals exceed the point source model 
(figure \ref{fig:gcsrc1} bottom) by a factor of $\sim5-7$. 
The effective solid angle of the XIS and PIN are $168$ and
1220 arcmin$^2$ respectively, and the factor is close to the solid angle difference
between the two as $1220/168=7.3$. 
We therefore presume that a significant fraction of the PIN signals comes from 
celestial sources which are extended to the same order as the HXD-PIN FOV.

In table~\ref{tab:pin_powerlaw_hxdnom}, 
some fits are not fully acceptable, and in such cases,  
the data are often found to be somewhat more convex than the model.
Therefore, we fitted the PIN spectra also with a cutoff power-law model,
\begin{equation}
f(E)=K(E/1~\mathrm{keV})^{-\Gamma}\exp({-E/E_c})\label{equ:cutoffpl},
\end{equation}
where $E$ is the energy, $f(E)$ is the photon flux per unit
energy interval, $K$ is a normalization factor in units of photons~keV$^{-1}$~cm$^{-2}$~s$^{-1}$, $\Gamma$ is again the photon index, 
and $E_{\rm c}$ represents a typical energy scale of the spectral cutoff.
The fit result improved slightly with reduced chi-squared of $0.75-1.5$,
and the derived photon index and cutoff energy varied
from region to region, over $1.2-2.2$ and $19-50$ keV, respectively. 

\begin{table}
\caption{Results of the fitting to the PIN spectra from the 5 regions with a single power-law model.}\label{tab:pin_powerlaw_hxdnom}
\begin{center}
\begin{tabular}{lccc}
\hline\hline
Name&Photon Index\footnotemark[$*$]&Flux\footnotemark[$\dagger$]&	$\chi^2_\nu$\footnotemark[$\ddagger$]\\
\hline
Region A$_1$ & $2.79\pm0.04$ & 2.86 & 0.98\\
Region A$_2$ & $2.68\pm0.04$ & 2.88 & 0.77\\
Region A$_3$ & $2.65\pm0.04$ & 2.66 & 1.42\\
Region B$_1$ & $2.62\pm0.04$ & 3.06 & 1.17\\
Region B$_2$ & $2.66\pm0.04$ & 2.89 & 1.39\\
Region C & $2.20\pm0.02$ & 6.28 & 1.69\\
Region D & $2.30\pm0.05$ & 1.53 & 1.02\\
Region E & $2.08\pm0.04$ & 3.35 & 1.41\\
\hline
\multicolumn{4}{@{}l@{}}{\hbox to 0pt{\parbox{180mm}{\footnotesize
\vspace{0.2cm}
\footnotemark[$*$]Errors are at 90\% confidence level.\par
\footnotemark[$\dagger$]$12-40~\mathrm{keV}$ flux in units of $10^{-10}~\ergcms$ \par
\footnotemark[$\ddagger$]With 72 degrees of freedom.
}\hss}}
\end{tabular}

\end{center}
\end{table}

\subsection{Bright point sources}
\label{subsec:brightpointsources}
In the three offset observations,
we clearly detected three bright point sources. 
As listed in table \ref{tab:obslog},
they have been identified with catalogued sources, 
namely 1A 1742$-$294, KS 1741$-$293 and 1E 1743.1$-$2843, 
which are well known through previous observations. 
During the present pointings for a few ks each,
these sources exhibited only mild variations up to 20\%.

In the following analysis, we need to
estimate the contribution of these point sources
to the PIN signals detected from Region A through E,
because the HXD has no imaging capability.
Our strategy is to quantify their spectra 
in the XIS energy band ($2-10$ keV),
and extrapolate the results to the  PIN band  ($12-40$ keV).
The PIN data obtained in the offset observations were
not used because they are contaminated by other
bright point sources and possibly by the Galactic center extended emission itself.
In order to minimize systematic errors in this process,
we also refer to the literature on their spectra,
obtained through past observations with various satellites.
In particular, we use the INTEGRAL light curves of the GC region \citep{kuu07}, 
which is extracted using OSA6 software and distributed 
by the INTEGRAL Science Data Center \citep{cou03},
to examine our spectral models for their consistency 
in the higher energy band ($20-60$ keV).
To reduce statistical errors, we averaged 
INTEGRAL IBIS count rate of each point source
over one week around our each observation.
The used spectral model and the best fit parameters obtained 
in this way are listed in table \ref{tab:psfit}, 
together with 90\% confidence errors.
Below, we briefly describe results on the three sources.

\begin{figure}
  \begin{center}
    \FigureFile(80mm,50mm){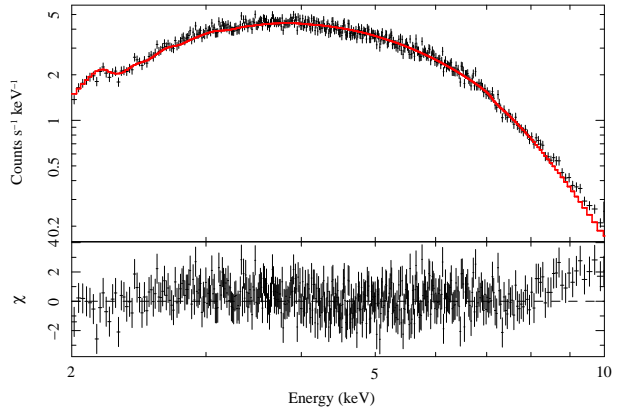}
  \end{center}
  \caption{The NXB-subtracted spectrum of Offset1$\_$1 (1A 1742$-$294) 
  taken with XIS FI CCDs (black crosses), 
  compared with the best fit bremsstrahlung model (red line). }\label{fig:offset}
\end{figure}

%1A 1742$-$294
1A 1742$-$294 is a persistent low-mass X-ray binary (LMXB),
and is known to harbor a neutron star,
since recursive type-I X-ray bursts have been observed (e.g., \cite{lew76}; \cite{sid99b}; \cite{sak02}).
\citet{sid99b} reported that 
an absorbed single power-law model with $\Gamma\sim1.7-1.9$,
or an absorbed thermal bremsstrahlung model 
with a temperature of $kT\sim10-16~\mathrm{keV}$,
gave an equally acceptable fits to the $2-10$ keV spectrum 
taken with the BeppoSAX MECS. 
Analyzing the XIS data, 
we obtained similar result with
somewhat softer spectral parameters;
a power law with $\Gamma\sim2.0-2.2$,
or a bremsstrahlung with $kT\sim8.0-9.4~\mathrm{keV}$.
The achieved fit result is presented in figure~\ref{fig:offset}.
When extrapolated to higher energies,
our bremsstrahlung model derived from the XIS data predicts
an intensity of $5.9-7.8$ mCrab\footnote{In the $20-60$ keV band, a 1 mCrab point source emits a flux of $1.38\times10^{-11}~\ergcms$.}
 in the $20-60$ keV INTEGRAL IBIS band.
This is in a good agreement with the INTEGRAL measurement \citep{kuu07},
that the $20-60$ keV  intensity of this source in the period of our observation 
was $\sim5\pm1$ mCrab. In contrast, the power-law model 
determined by the XIS data predicts a hard X-ray intensity
which is higher by a factor of $6-7$. 
We therefore regard the bremsstrahlung model as appropriate for the present purpose.

%KS 1741$-$293
KS 1741$-$293 is a transient LMXB with type-I X-ray bursts \citep{int91}. 
Its spectrum was so far described with an absorbed single power-law model 
or an absorbed bremsstrahlung model (\cite{sid99b}; \cite{int91}),
both yielding a relatively high hydrogen column density of 
$N_\mathrm{H}\sim10^{23} \mathrm{cm}^{-2}$. 
Applying these two models to the XIS data, 
we obtained a close  value as
$N_\mathrm{H}\sim(1.6-2.1)\times10^{23}\mathrm{cm}^{-2}$.
The photon index ($\Gamma\sim2.0-2.3$) of the power-law model 
and the temperature ($kT\sim7.7-10~\mathrm{keV}$) of the bremsstrahlung model, 
which we derived with the XIS,
are not much  different from the previous values of 
$\Gamma\sim2.0-2.1$ of \citet{sid99b} and 
$kT\sim9~\mathrm{keV}$ of \citet{int91}, respectively.
However, our bremsstrahlung modeling predicts a  
$20-60$ keV intensity of 0.7 mCrab, 
which is much lower than that measured with the INTEGRAL IBIS 
($\sim7\pm1$ mCrab).
Therefore, when estimating the contribution of this  source 
to the GC observations (see next subsection),
we instead use our power-law modeling,
which predicts  a $20-60$ keV intensity of 4.4 mCrab and 2.6 mCrab
for the first and second observations, respectively.

%1E 1743.1$-$2843
1E 1743.1$-$2843 is a persistent {X-ray source},
but the type of its compact object is unknown
because no type-I X-ray burst has been detected so far  (e.g., \cite{por03}; \cite{cre99}). 
BeppoSAX and ASCA observations showed
that its spectrum can be better described by
an absorbed blackbody model than by an absorbed power-law model 
or an absorbed bremsstrahlung \citep{cre99}. 
In fact, the XIS spectrum was reproduced 
successfully by a black body model with a temperature of $kT=1.8$ keV,
absorbed by $N_\mathrm{H}=13\times10^{22}$ cm$^{-2}$.
However, according to  \citet{del06}, 
wide-band spectra of this source,
taken simultaneously with XMM-Newton, Chandra,  and INTEGRAL IBIS,
require a steep power-law component with $\Gamma\sim3.1-3.3$
to be added  to a blackbody model of $kT\sim1.6-1.8~\mathrm{keV}$,
suggesting that the object may be a black-hole binary.
Since our purpose  is to extrapolate the spectrum from the XIS band to that of PIN, 
the hard tail component  should not be neglected.
Therefore, we also applied a black body plus 
power law model to the XIS data,
in which the $20-100$ keV  flux of the power-law component 
is fixed at $8\times10^{-12}$ erg s$^{-1}$ cm$^{-2}$
but its slope is allowed to vary.
This constraint on the flux was derived by scaling the $20-100$ keV flux of
$1.7\times10^{-11}$ erg s$^{-1}$ cm$^{-2}$ measured by \citet{del06},
to the flux ratio in the $2-10$ keV band, $\sim0.45$,
between their and our measurements.
We have then obtained $kT=1.8-1.9$ keV and $\Gamma=2.9-3.2$,
in a good agreement with  \citet{del06}.

\begin{figure}
  \begin{center}
    \FigureFile(80mm,50mm){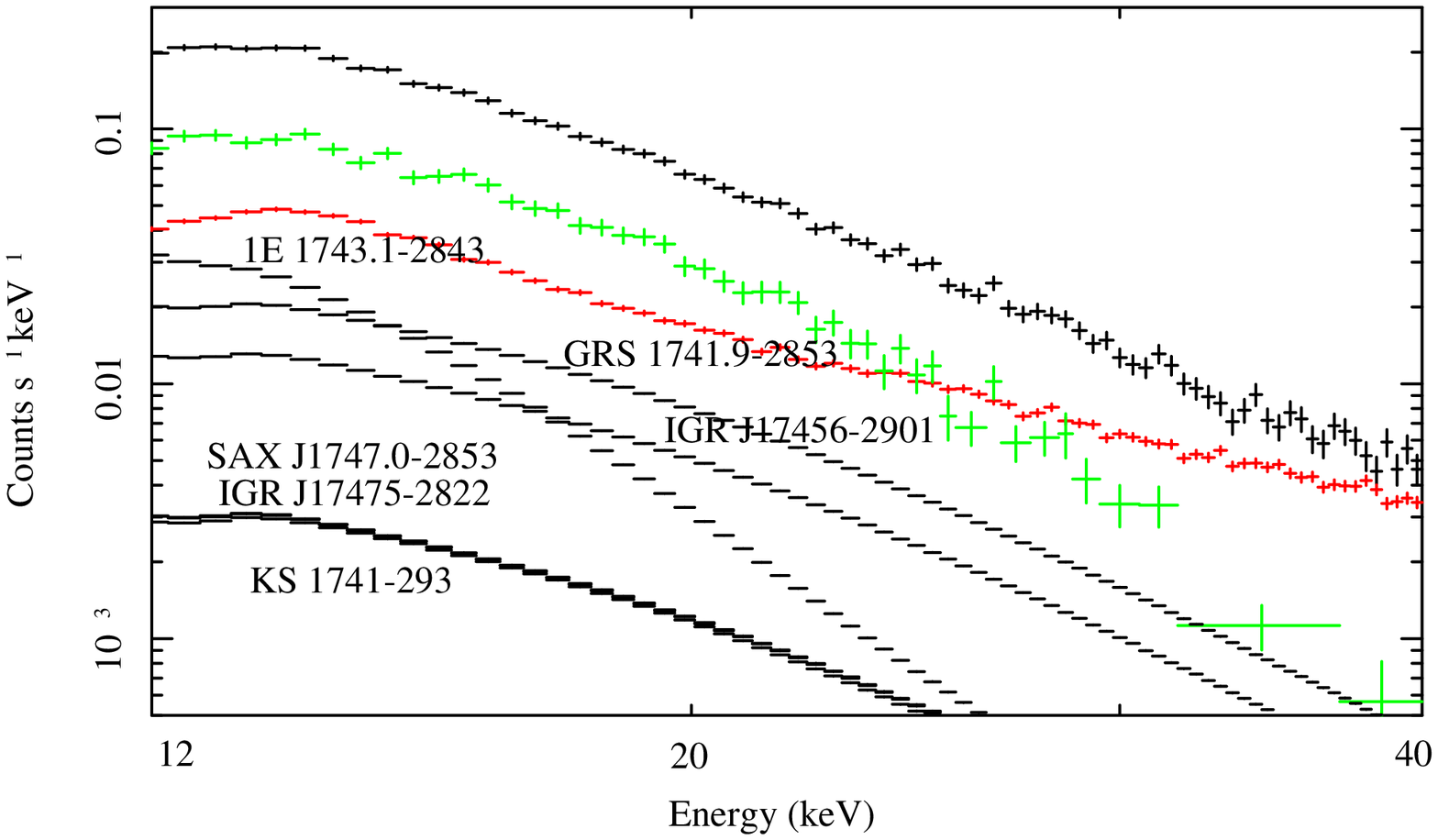}
    \FigureFile(80mm,50mm){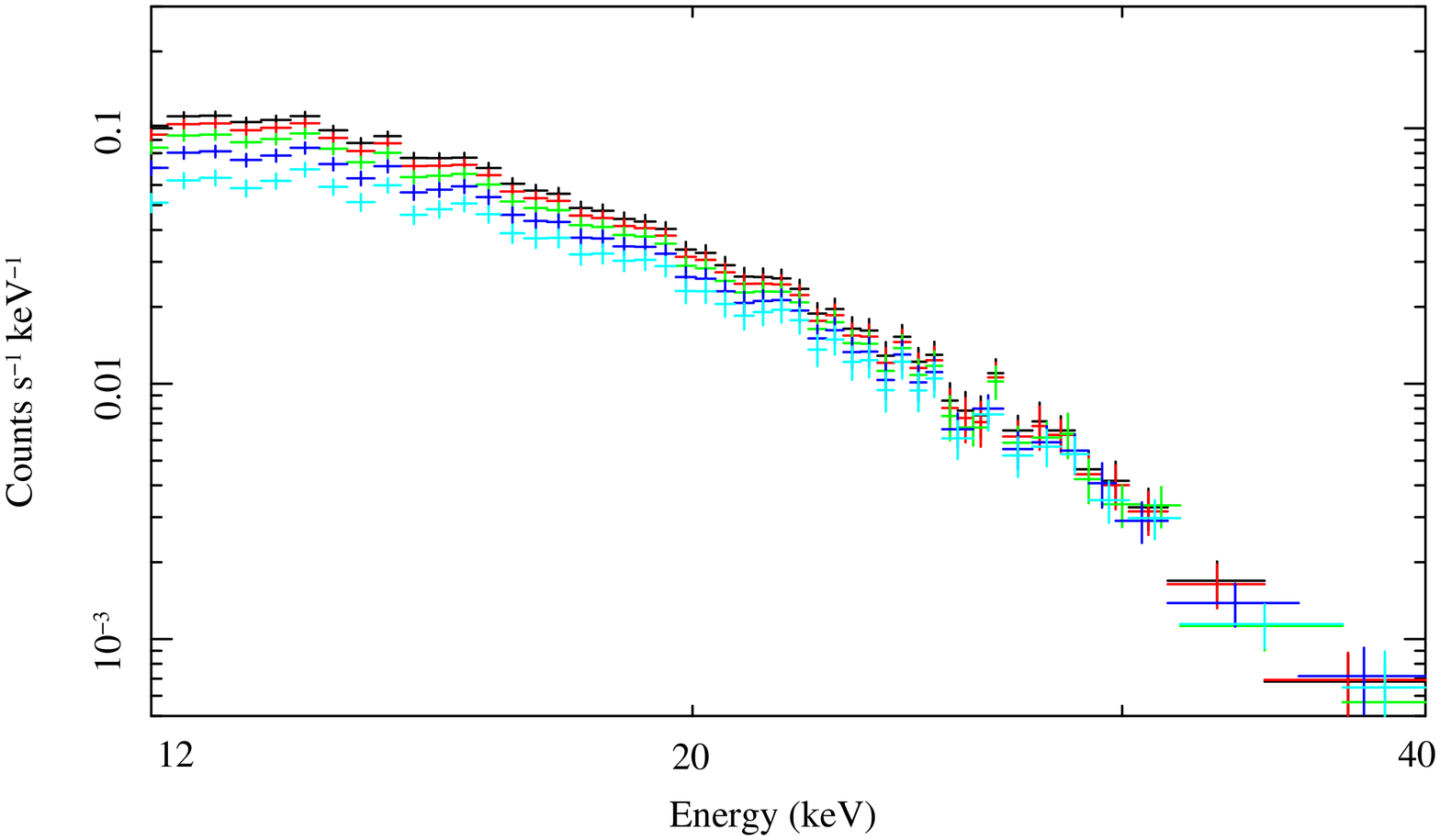}
  \end{center}
  \caption{Top : Raw PIN spectrum of Region A$_1$ (black), 
   compared with the modeled NXB spectrum (red). 
  %The simulated spectra of 1A 1742$-$294, KS 1741$-$293, and 1E 1743.1$-$2843 are plotted by cyanic, magenta, and yellow lines, while 
  Black lines show contributions from the bright point sources,
  estimated from the XIS offset observations for
  1A 1742$-$294, KS 1741$-$293, and 1E 1743.1$-$2843,
  or assuming power-law models of a photon index 
  $\Gamma_\mathrm{p}=2.1$ (for the ``other'' point sources). 
  The residual spectrum,
  derived after subtracting the NXB and the point-source contributions, 
  is plotted by green crosses. 
   Bottom : The same data as the green data points in the top panel, 
   but presented for four different point source estimations. 
   Red, green, blue, and cyanic crosses represent the residual spectra 
   obtained by changing $ \Gamma_{\rm p}$ of the ``other'' point sources
    in the top panel to 1.5, 1.8, 2.1, 2.4 and 2.7, respectively.}
  \label{fig:subtraction}
\end{figure}

%%%修正%%%
%PEGPL→PL%
%３カ所%
%%%%%%%%
\begin{table*}
\caption{Spectral parameters of the three point sources determined with the XIS offset observations.\footnotemark[$*$]}\label{tab:psfit}
\begin{center}
\begin{tabular}{llccccc}
\hline\hline
Name&Model\footnotemark[$\dagger$]&Column Density&\multicolumn{2}{c}{Parameters}&Flux\footnotemark[$\ddagger$]& $\chi^2_\nu$\\
&&$(10^{22}\mathrm{cm}^{-2})$&$kT(\mathrm{keV})$&$\Gamma$&& \\
\hline
Offset1$\_$1 (1A 1742-294)	&	TB	&	$6.0^{+0.1}_{-0.1}$	&	$8.0^{+0.3}_{-0.3}$	&	&	6.30	&	1.05 (1792)\\
	&	PL	&	$7.1^{+0.1}_{-0.1}$	&	&	$2.15^{+0.03}_{-0.03}$	&	6.35	&	1.02 (1792)\\
Offset1$\_$2 (1A 1742-294)	&	TB	&	$5.7^{+0.1}_{-0.1}$	&	$9.4^{+0.4}_{-0.3}$	&	&	5.64	&	1.13 (1915)\\
	&	PL	&	$6.7^{+0.1}_{-0.1}$	&	&	$2.03^{+0.03}_{-0.03}$	&	5.67	&	1.10 (1915)\\
Offset2$\_$1 (KS 1741-293)	&	TB	&	$18.9^{+1.1}_{-1.1}$	&	$7.6^{+1.4}_{-1.1}$	&	&	0.52	&	0.95 (319)\\
	&	PL	&	$20.9^{+1.5}_{-1.4}$	&	&	$2.3^{+0.1}_{-0.1}$	&	0.52	&	0.97 (319)\\
Offset2$\_$2 (KS 1741-293)	&	TB	&	$16.0^{+1.3}_{-1.2}$	&	$10.1^{+3.8}_{-2.1}$	&	&	0.23	&	0.92 (240)\\
	&	PL	&	$17.6^{+1.7}_{-1.6}$	&	&	$2.0^{+0.2}_{-0.2}$	&	0.23	&	0.92 (240)\\
Offset3$\_$1 (1E 1743.1-2843)	&	BB	&	$10.8^{+0.7}_{-0.7}$	&	$1.8^{+0.1}_{-0.1}$	&	&	1.03	&	1.00 (487)\\
	&	BB+PL	&	$15.0^{+2.4}_{-3.4}$	&	$1.9^{+0.1}_{-0.1}$	&	$3.0^{+0.3}_{-0.7}$	&	1.04	&	0.99 (486)\\	
Offset3$\_$2 (1E 1743.1-2843)	&	BB	&	$12.2^{+0.6}_{-0.6}$	&	$1.9^{+0.1}_{-0.1}$	&	&	1.60	&	1.10 (773)\\
	&	BB+PL	&	$17.4^{+2.4}_{-2.7}$	&	$2.0^{+0.1}_{-0.1}$	&	$3.2^{+0.2}_{-0.4}$	&	1.60	&	1.09 (772)\\
\hline
\multicolumn{7}{@{}l@{}}{\hbox to 0pt{\parbox{180mm}{\footnotesize
\vspace{0.2cm}
\footnotemark[$*$]Errors are at 90\% confidence level.\par
\footnotemark[$\dagger$]TB:thermal bremsstrahlung (\texttt{bremss} in \texttt{XSPEC}), PL:power law (\texttt{powerlaw}), BB:black body (\texttt{bbdoy}),
 BB+PL:black body \par
 plus power law with pegged flux $8\times10^{-12}\ergcms$ in $20-100$ keV. \par\noindent
\footnotemark[$\ddagger$]Unabsorbed flux ($2-10~\mathrm{keV}$) in units of $10^{-10}~\ergcms$\par
}\hss}}
\end{tabular}
\end{center}
\end{table*}

\begin{table}
\caption{The list of bright ($\gtrsim$1 mCrab in the $12-40$ keV band) point sources that can exist inside the PIN FOV in the observations of Region A$-$E.}\label{tab:pslist}
\begin{center}
\begin{tabular}{lcccc}
%%%%%%%%%%%
\hline\hline									
Name	&	$l$ ($^\circ$)	&	$b$ ($^\circ$)	&	\multicolumn{2}{c}{Rate (cts s$^{-1}$)\footnotemark[$*$]}			\\
	&	&	&			1st	&	2nd	\\
\hline									
1E 1740.7$-$2942	&	-0.88	&	-0.11	&	$10.2$	&	$10.4$	\\
1A 1742$-$294\footnotemark[$\dagger$]	&	-0.44	&	-0.39	&	$1.2$	&	$1.3$	\\
KS 1741$-$293\footnotemark[$\ddagger$]	&	-0.43	&	-0.09	&	$1.9$	&	$1.7$	\\
IGR J17456$-$2901	&	-0.06	&	-0.05	&	$0.7$	&	$0.7$	\\
GRS 1741.9$-$2853	&	-0.05	&	0.12	&	$1.3$	&	$0.5$	\\
SAX J1747.0$-$2853	&	0.21	&	-0.24	&	$0.2$	&	$0.2$	\\
1E 1743.1$-$2843\footnotemark[$\S$]	&	0.26	&	-0.03	&	$1.2$	&	$1.0$	\\
IGR J17475$-$2822	&	0.61	&	-0.06	&	$0.8$	&	$0.8$	\\
IGR J17497$-$2821	&	0.95	&	-0.45	&	$1.0$	&	$0.9$	\\
\hline
%%%%%%%%%%%
\multicolumn{5}{@{}l@{}}{\hbox to 0pt{\parbox{80mm}{\footnotesize
\vspace{0.2cm}
\footnotemark[$*$]A week averaged INTEGRAL IBIS count rate ($20-60$ keV) in the first and second observation of Region A and B with a typical error of $\pm0.3$ cts s$^{-1}$. 5 cts s$^{-1}$ corresponds to $\sim$20 mCrab ($2.8\times 10^{-10}~\ergcms$) in the $20-60$ keV band.\par
\footnotemark[$\dagger$]Offset1 \footnotemark[$\ddagger$]Offset2 \footnotemark[$\S$]Offset3\par
}\hss}}
\end{tabular}
\end{center}
\end{table}

\subsection{Subtraction of bright point source contributions from the PIN spectra}
\label{subsec:subtraction}
As can be expected from figure \ref{fig:gcimage}, 
the PIN signals obtained in the present observations should be mixtures of 
the extended emission around the GC and the emission from known bright point sources. 
Therefore, to derive the surface brightness 
and net spectra of the extended emission, 
we need to estimate the contribution of those point sources and subtract it from the PIN spectra. 
For this purpose, we selected catalogued point sources 
that can be brighter than 1 mCrab in our observations, 
and located in the $l<|\timeform{1\circ.5}|$ and $b<|1^\circ|$ region 
because the full FOV of PIN is $\sim1^\circ\times1^\circ$. 
At the 8 kpc distance, the threshold of 1 mCrab corresponds approximately
to a $2-40$ keV luminosity of $\sim3.5\times10^{35}~\mathrm{erg}~\mathrm{s}^{-1}$.
Table \ref{tab:pslist} shows a list of the selected sources ordered by their Galactic longitudes.
Although the PIN signals must be contributed also by more numerous 
uncatalogued (and mostly dimmer) point sources, 
we include them into what we call ``extended'' emission, because, in hard X-ray band, 
their contribution cannot be estimated at present with a sufficient reliability.

Here, we limit our detailed analysis to the first and second observations of Region A and B, 
namely A$_1$, A$_2$, B$_1$, and B$_2$,
because the XIS offset observations of the three point sources 
were conducted within a few days of these four pointings, 
and hence their contributions can be most accurately estimated.
Specifically, we estimated the $12-40$ keV spectra of the three objects, 
by extrapolating the best fit model obtained in \S\ref{subsec:brightpointsources}; 
a bremsstrahlung with $kT=7.98$ and 9.37 keV for 1A 1742$-$294, 
a power law with $\Gamma=2.26$ and 2.04 for KS 1741$-$293, 
and a blackbody ($kT=1.85, 1.98$ keV) plus power law ($\Gamma=2.95, 3.21$) 
for 1E 1743.1$-$2843. 
Each parameter is assigned a pair of values,
applicable in this order to the first and second observations of those objects. 
The estimated spectra were then multiplied by 
the angular response of HXD-PIN (figure 8 of \cite{tak07}), 
considering angular offsets of these objects 
(from the PIN optical axis) in each pointing observation.
The relative normalization difference by $\sim13\%$, between the XIS and PIN 
\citep{kok07}, was also considered by multiplying 1.13 to the normalization
parameter derived from the XIS spectra.

Besides the three sources, 
there are 6 catalogued point sources, as listed in table \ref{tab:pslist}, 
which can contribute to the PIN signals. 
Since we do not have direct XIS information on them, 
we assumed all of them to have power-law spectra 
with a common photon index $\Gamma_\mathrm{p}$, 
and adjusted individual normalizations 
so that their week-averaged $20-60$ keV intensities, 
as recorded by INTEGRAL IBIS (\S\ref{subsec:brightpointsources}), 
can be reproduced. 
Their spectra estimated in this way were then multiplied by the HXD-PIN angular response, 
in the same way as for the preceding three sources.
 
Figure \ref{fig:subtraction} (top panel) exemplifies 
the estimated spectra of the listed point sources, 
and the residual PIN spectra (green) obtained after 
subtracting the NXB and all these point source contributions.
Since the current PIN response below 12 keV contains rather large uncertainty,
the point source simulation was conducted in the $12-40$ keV band.
The bottom panel of figure \ref{fig:subtraction} illustrates 
how the residual PIN spectrum changes as the common photon index 
assumed for the point sources (other than the three) is varied 
as $\Gamma_\mathrm{p}=1.5,~1.8,~2.1,~2.4$ and 2.7. 
Thus, the residual PIN signal is highly significant, 
even $\Gamma_\mathrm{p}$ is changed over a plausible range. 
The difference of the residual flux between the two extreme cases,
$\Gamma_\mathrm{p}=1.5$ and 2.7, is $\sim2$.

We analyzed the data from Region A$_2$, B$_1$, and B$_2$ 
in the same manner as Region A$_1$. 
The obtained residual PIN spectra are shown in figure \ref{fig:regionab_residual_spectra}, 
where contributions of the three point sources were treated individually, 
while those of the other listed sources were modeled 
with $\Gamma_\mathrm{p}=2.1$ and adjusted as described before.
Thus, the residual signals from Region A$_2$, B$_1$ and B$_2$ are also significant. 
Furthermore, the signals from Region A and Region B are 
very similar both in the normalization and the spectral slope, 
even after removing the estimated contributions from the known point sources.

As mentioned in \S\ref{subsection:spectra}, 
the ratio between the XIS and PIN signals are different from 
what is expected from a point source. 
We roughly explained the PIN signal excess over 
a predicted point source spectrum, based on its larger effective solid angle than that of the XIS, and the uniform spatial extent of the emission. 
Then, we subtracted the bright point source contribution, and found the PIN signals to be roughly halved in each region.
Nevertheless, these residual PIN spectra still exceed significantly the model spectrum for a 5 mCrab point source (blue solid line in figure \ref{fig:gcsrc1} bottom),
reconfirming the extended nature of the emission as noted in \S\ref{subsection:spectra}. We further consider the surface brightness distribution of the extended emission in \S\ref{subsection:sifit}.

\subsection{Fitting to the HXD spectra}
\label{subsection:HXDfits}
\begin{figure}
  \begin{center}
    \FigureFile(80mm,50mm){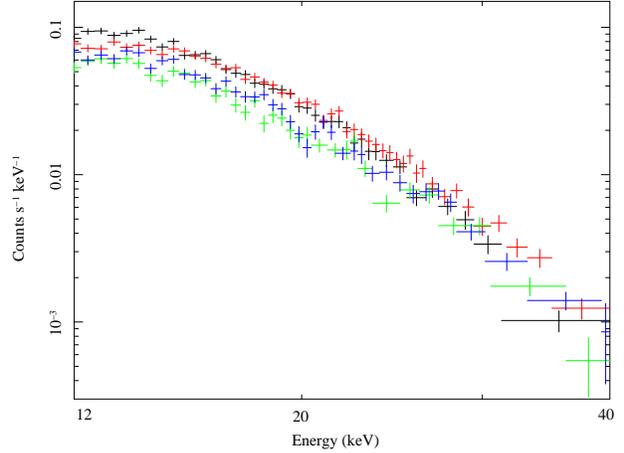}
  \end{center}
  \caption{The residual PIN spectra of Region A$_1$ (black), A$_2$ (red), 
  B$_1$ (green), and B$_2$ (blue). 
  Contributions from 1A 1742$-$294, KS 1741$-$293 and 1E 1743.1$-$2843 
  were estimated and subtracted individually (see text), 
  while those from the other catalogued point sources were removed 
  assuming $\Gamma_\mathrm{p}=2.1$.}
  \label{fig:regionab_residual_spectra}
\end{figure}

Assuming that the emission uniformly fills the HXD-PIN FOV, 
we fitted the residual PIN spectra (figure \ref{fig:regionab_residual_spectra})
 with several simple models; 
a thermal bremsstrahlung, 
a power law, a broken power law, and a cutoff power law of equation (\ref{equ:cutoffpl}).
A broken power-law model is expressed as
\begin{eqnarray}
f(E)=
\left
\{
    \begin{array}{cc}
        K(E/1~\mathrm{keV})^{-\Gamma_1}, & E<E_c\\
        KE_c^{\Gamma_2-\Gamma_1}(E/1~\mathrm{keV})^{-\Gamma_2}, & E>E_c,\\
    \end{array}
\right.
\end{eqnarray}
where $K$ is a normalization factor similar to that in equation (\ref{equ:cutoffpl}), and $E_c$ is an energy of a spectral break point. The photon indices below and above $E_c$ are denoted as $\Gamma_1$ and $\Gamma_2$, respectively.
To avoid the uncertainty included in the PIN response below 12 keV, we limited the PIN spectral fitting to the $12-40$ keV band.
The model spectrum contains the small contribution 
of the cosmic X-ray background, which is typically $\sim5$~\% of the NXB. 
Results of these fits are summarized in
table \ref{tab:residual_spectra_of_regionab_fit},
together with the $12-40$ keV surface brightness implied by the model.

\begin{figure}
  \begin{center}
    \FigureFile(80mm,50mm){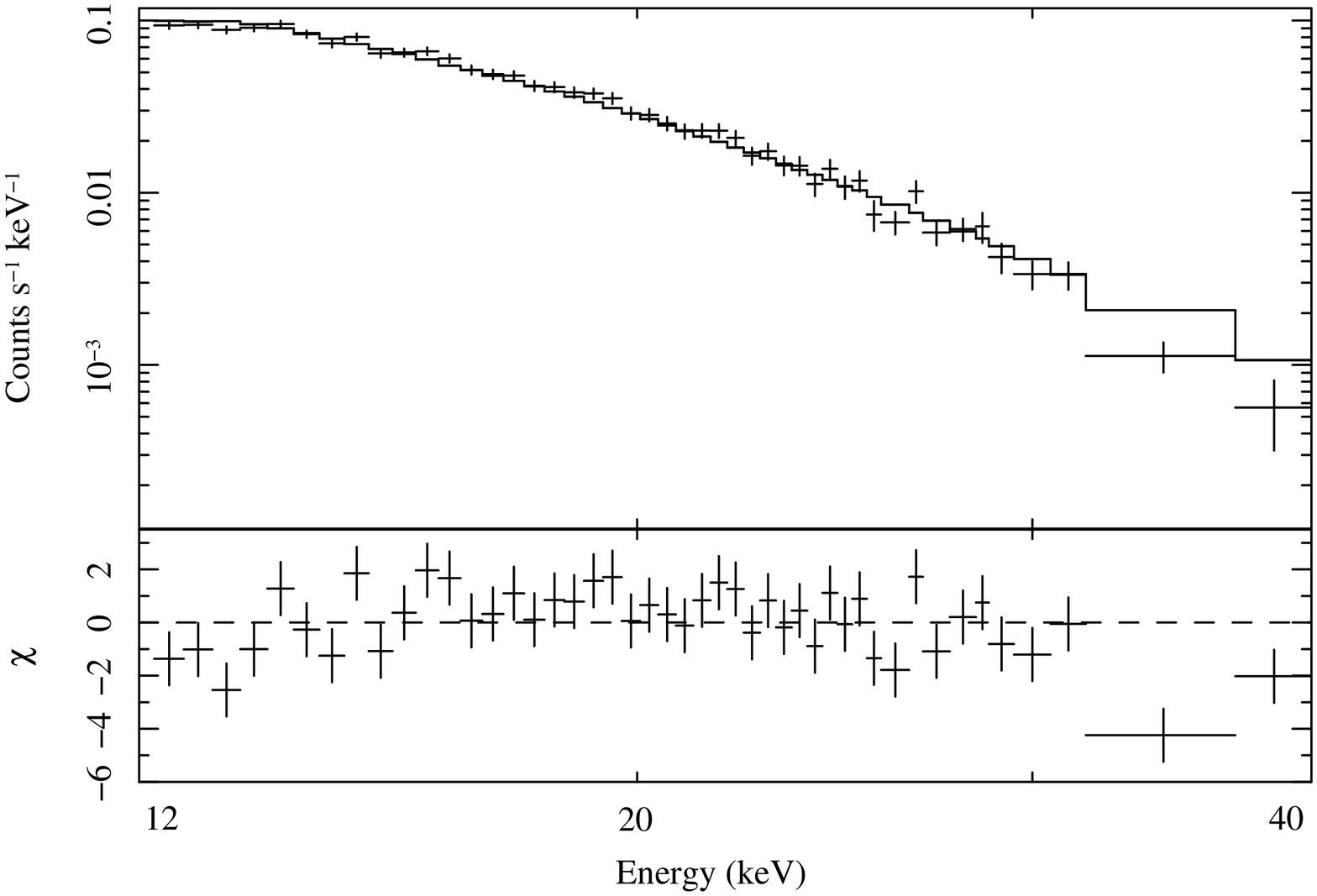}
    \FigureFile(80mm,50mm){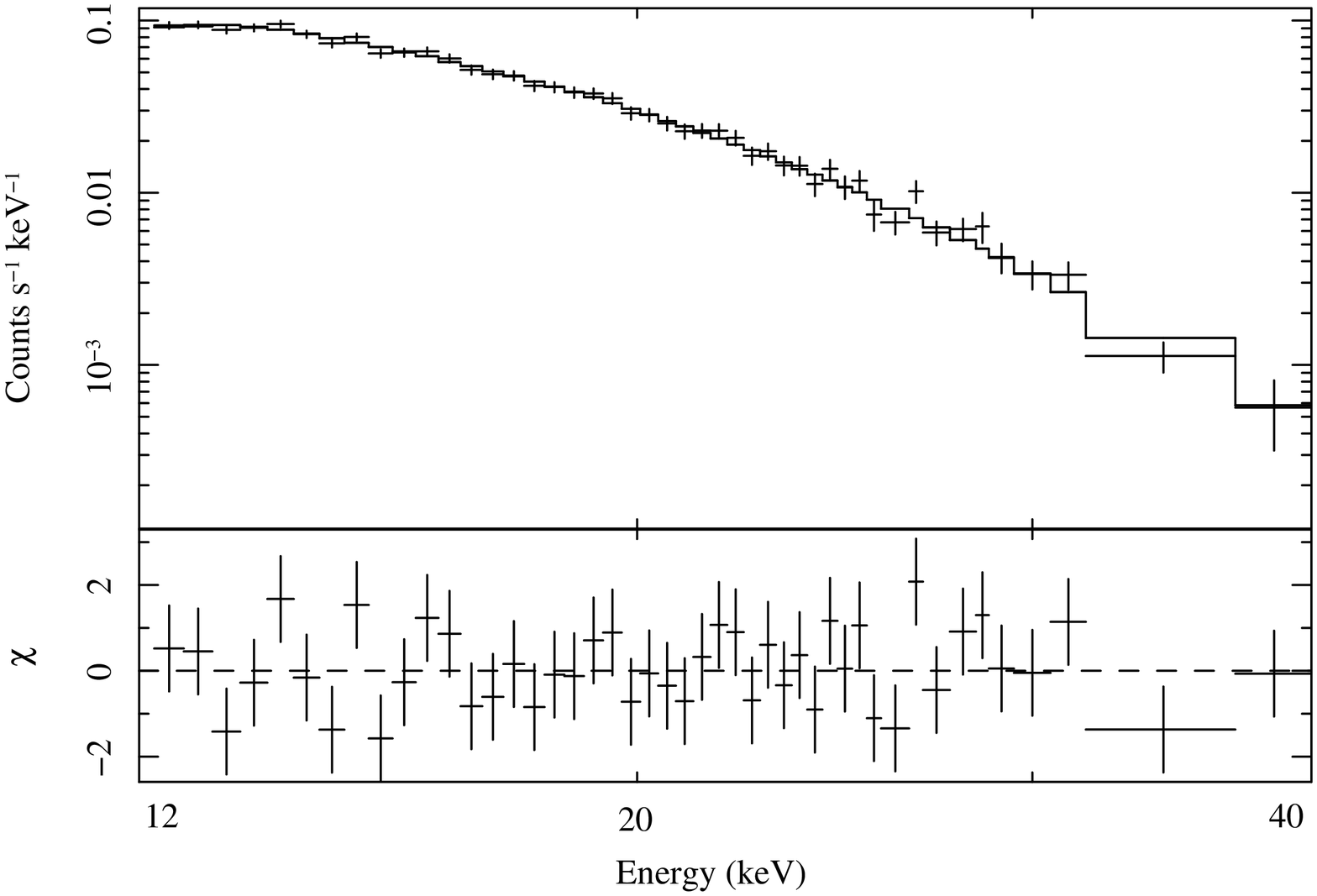}
  \end{center}
  \caption{The residual PIN spectrum of Region A$_1$ (black crosses in figure \ref{fig:regionab_residual_spectra}) fitted with a power-law model (top) and a cutoff power-law model (bottom).}
  \label{fig:residual_spectra_of_regionab_fit}
\end{figure}

Table~\ref{tab:residual_spectra_of_regionab_fit} indicates
that the residual spectrum of Region A (both A$_1$ and A$_2$)
prefer convex models (thermal bremsstrahalung, broken power law,
and cutoff power law) to the straight power-law model.
Indeed, as shown in figure \ref{fig:residual_spectra_of_regionab_fit},
a power-law fit to the Region A$_1$ spectrum leaves
significant fit residuals in the $<15$ keV and $>30$ keV regions,
while a cutoff power-law model is fully acceptable.
This is the same tendency as already noticed at the end of
\S\ref{subsection:spectra} before subtracting the point source contributions.
Spectra of Region B exhibit a similar preference,
though less significant.
If, e.g., the thermal bremsstrahlung fits are employed,
the spectra from all these regions are characterized
by a very high temperature of $\sim 15$ keV.
The cutoff power-law fits lead to similar cutoff ``temperature''.
If, instead, the broken power-law modeling is adopted,
the four spectra are consistently represented
by a photon index of $\Gamma_1 \sim 2$ and $\Gamma_2 \sim 3$
in energies below and above $\sim 20$ keV, respectively.

We examined how the fit results are affected
by the value of $\Gamma_\mathrm{p}$ assumed for the ``other'' point sources.
As a representative case,
the bottom half of table \ref{tab:residual_spectra_of_regionab_fit}
gives the best fit parameters of Region A$_1$,
obtained by changing $\Gamma_\mathrm{p}$ from 2.1 to 1.8, or to 2.4.
Since the $20-60$ keV fluxes of the ``other'' point sources
are individually fixed using the INTEGRAL data,
a harder (smaller) value of $\Gamma_\mathrm{p}$ makes
the residual PIN spectrum softer.
Nevertheless,
the implied $12-40$ keV surface brightness does not differ by more than  $\sim20\%$ 
between the cases with $\Gamma_\mathrm{p}=1.8$ and $\Gamma_\mathrm{p}=2.4$.

\begin{table*}
\caption{Results of model fits to the residual PIN spectra.\footnotemark[$*$]}
\label{tab:residual_spectra_of_regionab_fit}
\begin{center}
\begin{tabular}{clcccccc}
\hline\hline
Region	&	Model\footnotemark[$\dagger$]	&	\multicolumn{4}{c}{Parameters}	&	$\Sigma$\footnotemark[$\S$]	&	$\chi^2_\nu~(\nu)$\\
	&		&	$kT$ (keV)	&	$\Gamma_1$	&	$\Gamma_2$	&	$E_\mathrm{c}$\footnotemark[$\ddagger$] (keV)	&		&	\\
\hline
A1	&	TB	&	$12.8^{+0.8}_{-0.7}$	&		&		&		&	4.70	&	0.86 (72)\\
	&	BKNPL	&		&	$2.5^{+0.0.2}_{-0.2}$	&	$4.1^{+1.6}_{-0.6}$	&	$21.9^{+4.3}_{-3.1}$	&	4.63	&	0.87 (70)\\
	&	CUTOFFPL	&		&	$0.9^{+0.5}_{-0.7}$	&		&	$9.3^{+2.9}_{-2.5}$	&	4.65	&	0.83 (71)\\
	&	PL	&		&	$2.8^{+0.1}_{-0.1}$	&		&		&	4.85	&	1.38 (72)\\
B1	&	TB	&	$16.3^{+1.9}_{-1.6}$	&		&		&		&	3.28	&	1.20 (72)\\
	&	BKNPL	&		&	$2.5^{+0.1}_{-0.1}$	&	$27.0^{+0.1}_{-23}$	&	$36.7^{+2.0}_{-7.7}$	&	3.30	&	1.13 (70)\\
	&	CUTOFFPL	&		&	$2.0^{+0.5}_{-0.7}$	&		&	$35.4^{+164}_{-35.3}$	&	3.33	&	1.18 (71)\\
	&	PL	&		&	$2.6^{+0.1}_{-0.1}$	&		&		&	3.35	&	1.18 (72)\\
B2	&	TB	&	$17.4^{+1.7}_{-1.5}$	&		&		&		&	3.78	&	1.33 (72)\\
	&	BKNPL	&		&	$1.6^{+2.9}_{-2.4}$	&	$2.6^{+0.1}_{-0.1}$	&	$14.0^{+4.8}_{-1.2}$	&	3.83	&	1.30 (70)\\
	&	CUTOFFPL	&		&	$1.8^{+0.7}_{-0.6}$	&		&	$27.1^{+172.9}_{-12.4}$	&	3.80	&	1.33 (71)\\
	&	PL	&		&	$2.5^{+0.1}_{-0.1}$	&		&		&	3.88	&	1.35 (72)\\
A2	&	TB	&	$21.9^{+1.8}_{-1.6}$	&		&		&		&	5.20	&	0.85 (72)\\
	&	BKNPL	&		&	$1.3^{+0.4}_{-0.9}$	&	$2.7^{+0.2}_{-0.2}$	&	$16.6^{+2.0}_{-1.9}$	&	5.15	&	0.67 (70)\\
	&	CUTOFFPL	&		&	$0.3^{+0.4}_{-0.6}$	&		&	$10.1^{+3.1}_{-2.5}$	&	5.10	&	0.69 (71)\\
	&	PL	&		&	$2.3^{+0.1}_{-0.1}$	&		&		&	5.30	&	1.35 (72)\\
\hline														
A$_1$ ($\Gamma_\mathrm{p}=1.8$)\footnotemark[$\|$]	&	TB	&	$12.2^{+0.7}_{-0.6}$	&		&		&		&	5.08	&	0.87 (72)\\
	&	BKNPL	&		&	$2.6^{+0.1}_{-0.2}$	&	$4.2^{+1.4}_{-0.6}$	&	$22.0^{+3.8}_{-2.8}$	&	5.00	&	0.88 (70)\\
	&	CUTOFFPL	&		&	$0.9^{+0.5}_{-0.7}$	&		&	$9.0^{+2.7}_{-2.3}$	&	5.03	&	0.84 (71)\\
	&	PL	&		&	$2.9^{+0.1}_{-0.1}$	&		&		&	5.25	&	1.50 (72)\\
A$_1$ ($\Gamma_\mathrm{p}=2.4$)\footnotemark[$\|$]	&	TB	&	$14.1^{+1.0}_{-1.0}$	&		&		&		&	4.30	&	0.86 (72)\\
	&	BKNPL	&		&	$2.4^{+0.2}_{-0.2}$	&	$3.9^{+1.7}_{-0.6}$	&	$21.7^{+4.9}_{-3.4}$	&	4.23	&	0.86 (70)\\
	&	CUTOFFPL	&		&	$0.8^{+0.5}_{-0.8}$	&		&	$9.4^{+3.6}_{-2.7}$	&	4.25	&	0.82 (71)\\
	&	PL	&		&	$2.7^{+0.1}_{-0.1}$	&		&		&	4.43	&	1.28 (72)\\	
\hline
\multicolumn{8}{@{}l@{}}{\hbox to 0pt{\parbox{130mm}{\footnotesize
\vspace{0.2cm}
\footnotemark[$*$]Errors are at 90\% confidence level.\par
\footnotemark[$\dagger$]TB : thermal bremsstrahlung (\texttt{bremss} in \texttt{XSPEC}). BKNPL : broken power law (\texttt{bknpower}). CUTPL : cutoff power law (\texttt{cutoffpl}). PL : power law (\texttt{powerlaw}).\par
\footnotemark[$\ddagger$]The cutoff energy in the cutoff power-law model or the energy for the broken point in the broken power-law model.\par
\footnotemark[$\S$]$12-40~\mathrm{keV}$ surface brightness in units of $10^{-10}~\ergcms~\mathrm{deg}^{-2}$ \par
\footnotemark[$\|$]Results obtained by changing the photon index $\Gamma_\mathrm{p}$ for ``other'' point sources to 1.8 or 2.4.\par
}\hss}}
\end{tabular}
\end{center}
\end{table*}

\subsection{Simultaneous fitting to the XIS and PIN spectra}
\label{subsection:sifit}
Analyzing the XIS data of the extended GC emission, 
\citet{koy07a} reported the presence of a power-law-like 
hard tail with a photon index of $1.4^{+0.5}_{-0.7}$,
in addition to  the line-rich hot ($\sim6.5$ keV) thermal emission. 
However, in the XIS range, 
these two components have  relatively similar spectral slopes. 
Therefore, to more accurately distinguish them,
it is important to expand the available energy range by combining the XIS and HXD results. 

The residual PIN spectra of Region A and B,
obtained in the previous subsection by excluding 
the contribution from the catalogued bright point sources,
have a $10-15$ keV slope of $\sim 1.5$, 
in terms of a  broken power-law modeling.
Since this is close to that of  the XIS hard tail found by \citet{koy07a}, 
the residual PIN spectra are considered  to reflect mainly 
the extended GC emission (its hard tail component in particular)
that is also observed by the XIS.
Therefore, we tried a simultaneous fit to the XIS and PIN data.
Hereafter, we set $\Gamma_\mathrm{p}=2.1$  to subtract the 
contributions of the ``other'' point sources from the PIN signals.
 
One obvious problem in the XIS plus HXD simultaneous fitting is
that the HXD, with a larger FOV,
receives the extended signals even from outside the XIS FOV.
However, according to \citet{koy07a} and \citet{nob07},
the [Fe\emissiontype{XXVI} K$\alpha$]/[Fe\emissiontype{XXV} K$\alpha$] line flux ratio
of the GC extended emission in the $\timeform{-0\circ.4}<l<\timeform{0\circ.6}$ region
is almost constant at $\sim0.4$.
Therefore, we can assume, at least in Region A,
that  the spectral shape is constant inside the PIN FOV,
and only the surface brightness varies from place to place.
Since the Fe line flux ratio has not yet been precisely measured in $l<\timeform{-0\circ.5}$, 
and the PIN FOV during the Region B observation
extends much into the $l<\timeform{-0\circ.5}$ region,
below we limit our analysis to the Region A data.
Then, we combined the XIS and PIN spectra from the two observations
(A$_1$ and A$_2$) and derived the averaged spectra,
because the two observations  gave very similar spectra (figure \ref{fig:gcsrc1} and \ref{fig:gc5regions}).
To avoid  contributions of the cooler ($\lesssim1$~keV) thermal component,
we limited the fit energy range of the XIS data to $>5.5$ keV after  \citet{koy07a}.

When using the XIS data from Region A,
another particular caution is needed;
we must exclude signals from the bright Sgr A East region
(or the Sgr A$^*$ complex).
According to \citet{koy07c},
the XIS spectrum of this region exhibits two thermal components,
and  a power-law hard tail with $\Gamma\sim0.8$
which is estimated to contribute a flux of
$5.2\times10^{-11}~\ergcms$
to the $12-40$ keV  PIN data
(before applying the $65-75\%$ PIN  angular transmission).
We then excluded this region from the XIS event integration,
by masking a circular region of radius $3\prime$ centered on $(l,b)=(\timeform{+359\circ.95},\ \timeform{-0\circ.050})$.

When we turn to the PIN data, in \S\ref{subsec:subtraction}
we have subtracted the contribution from ``point source'' IGR J17456-2901,
which is reported to coincide with Sgr A East region and Sgr A* (e.g., \cite{ner05}).
During our observations of Region A and B,
IGR J17456-2901 was reported to exhibit roughly a constant $20-60$ keV flux of  
$(4.1\pm1.7)\times10^{-11}~\ergcms$ by INTEGRAL IBIS,
which predicts the $12-40$ keV flux of
%4.1 \ergcms (20-60 keV) = 2.971 mCrab
%2.971mCrab(12-40)=2.971*1.6\times10^{^11}=4,75 \ergcms
$(4.8\pm2.0)\times10^{-11}~\ergcms$
when extrapolated with  $\Gamma=2.1$. 
Since this value agrees well with that of the excluded XIS region,
and the thermal component of the emission decreases rapidly 
becoming much less dominant than 
the power-law component in the PIN band, 
we conclude that the contribution from the Sgr A East region
has been removed from both the XIS and HXD data in a consistent manner.

In the combined fitting, we used the following three models;
(a) a collisional ionization equilibrium (CIE) plasma emission (\texttt{apec} version 1.3.1) 
plus three gaussian lines,
(b) the same as (a) but a power law is added; and
(c) the same as (b) but the power law is replaced by a cutoff power-law model.
Among them, (b) is the same as in \citet{koy07a}.
The three gaussians represent the neutral (or low ionizaed) 
Fe K$\alpha$, Fe K$\beta$, and Ni K$\alpha$ lines.
In each model, we subjected all the model components to 
a common absorption fixed at $6 \times 10^{22}$ cm$^{-2}$,
and multiplied the PIN model with a constant (but free) factor;
the latter is intended to compensate for the difference 
of the model normalization between the two instruments, 
caused by possible nonuniformity of the surface brightness:
an implicit assumption is that the multiple spectral components
comprising a model have similar surface brightness distributions
within each PIN FOV. 
The iron abundance of the \texttt{apec} model 
was at first allowed to vary freely, 
but was not well constrained.
Therefore we fixed it to 1.0 solar, 
and examined the result by changing it as 0.5 and 2.0 (see below).  
Additionally, we introduced a small red-shift to the XIS model spectrum,
to compensate for the uncertainty of gain calibration as noted in \citet{koy07a}.
The backside illuminated CCD chip (XIS1) is not utilized in the combined fit because
it suffers from rather high NXB counts above $\sim7-8$ keV. We assumed
a circular emission region with a radius of $\timeform{60'}$ in calculating
the XIS ancillary response file by \texttt{xissimarfgen}.
With these assumptions, we expect the XIS versus PIN normalization ratio
to become unity when the emission has a uniform brightness over
the detector FOVs.

\begin{figure*}
  \begin{center}
    \FigureFile(80mm,50mm){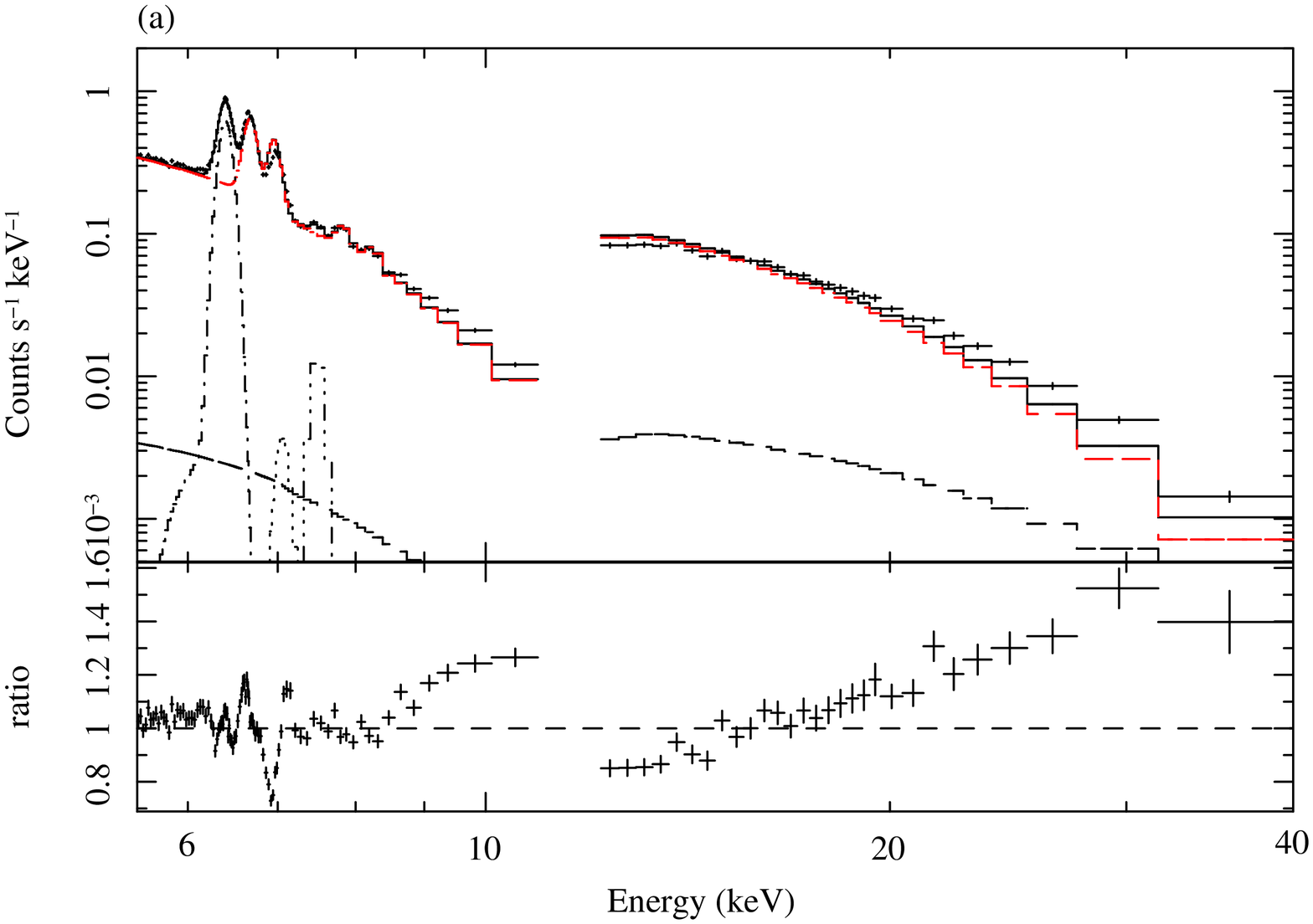} 
    \FigureFile(80mm,50mm){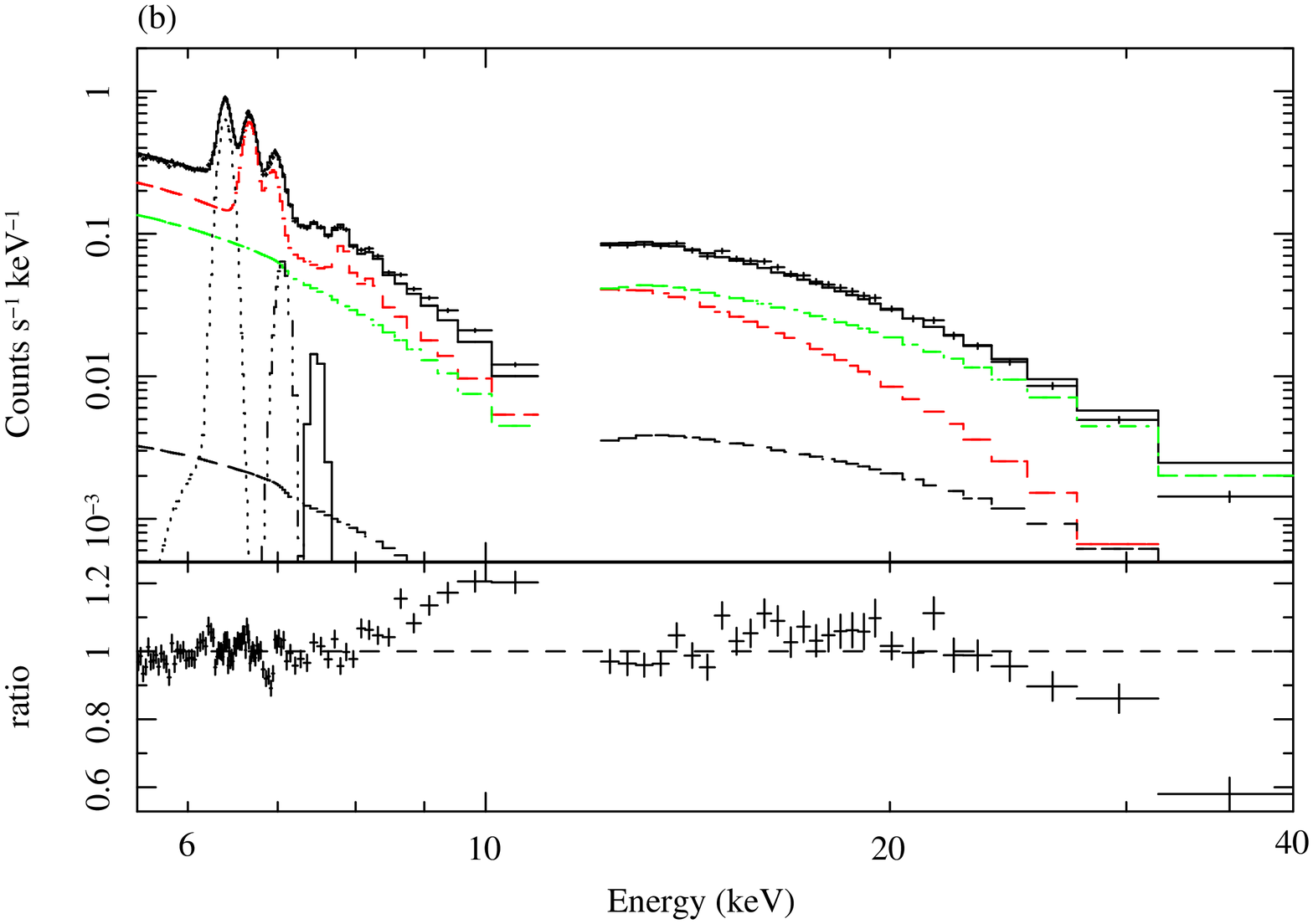}
    \FigureFile(80mm,50mm){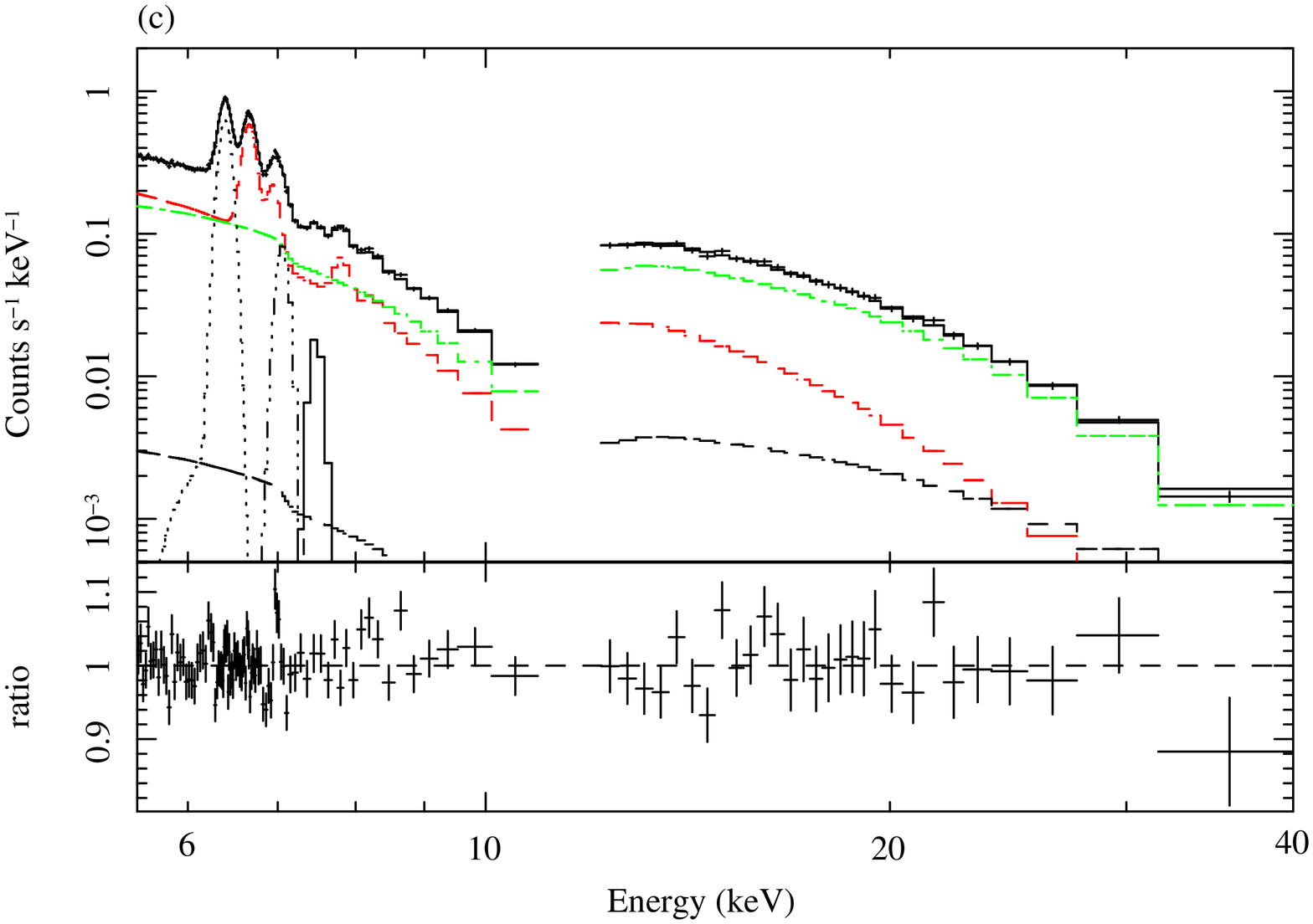}
    \FigureFile(80mm,50mm){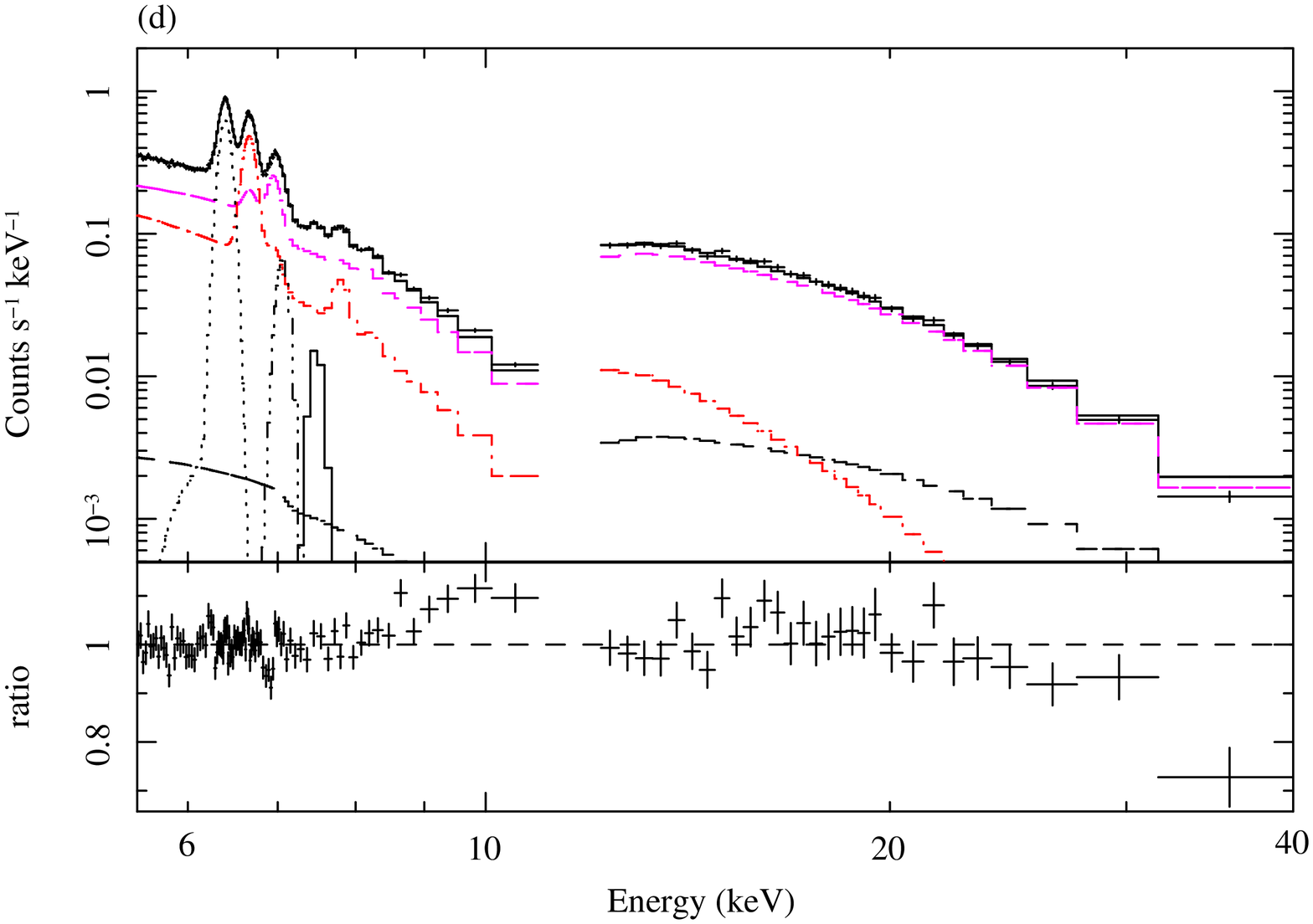}
    \FigureFile(80mm,50mm){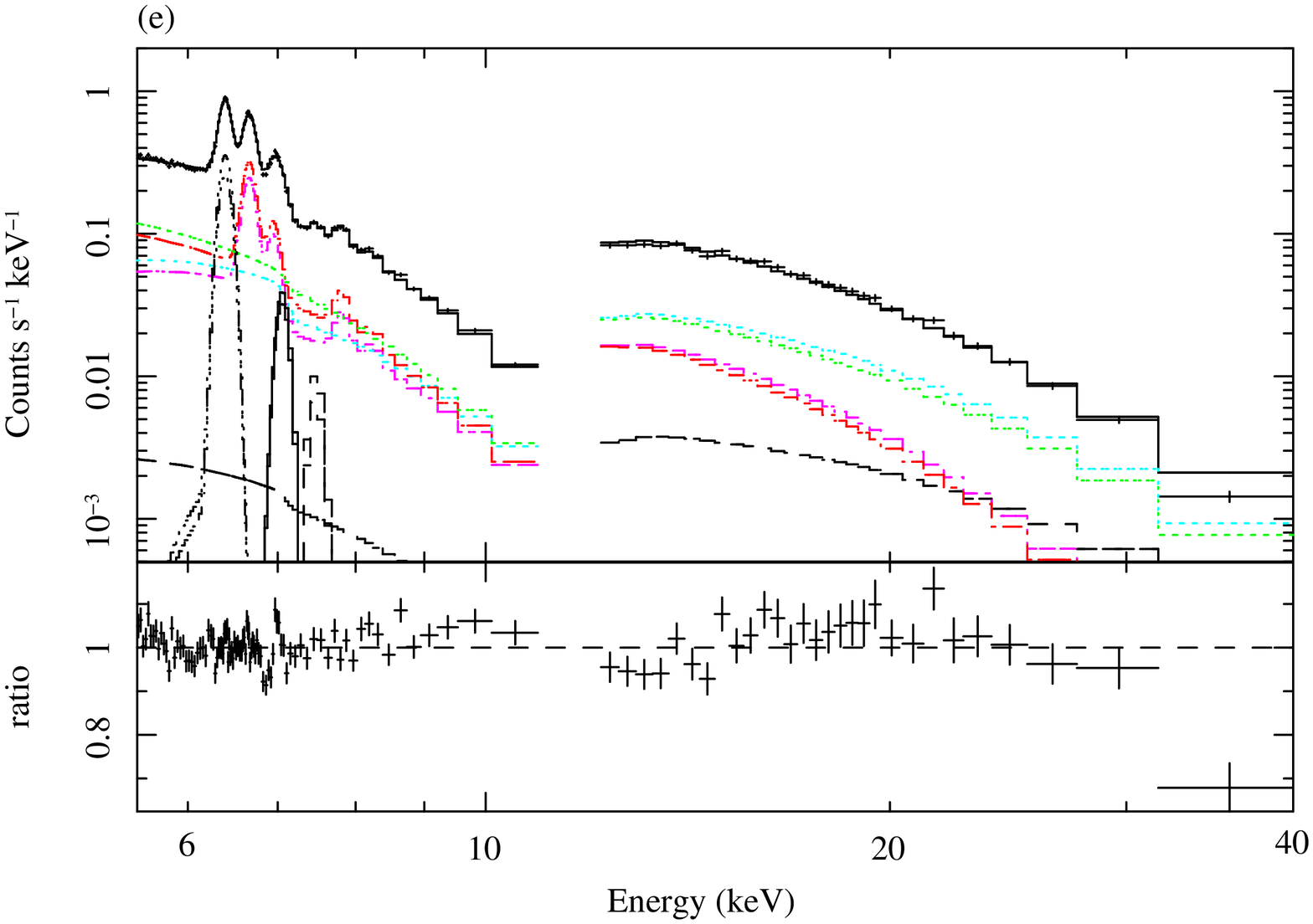}
  \end{center}
  \caption{The XIS and PIN spectra from Region A
 averaged over the first and second observations, 
 simultaneously fitted by different models.
 The cosmic X-ray background is included as a fixed model, which is indicated by black dashed lines in each panel.
The fit residuals are plotted in the bottom half 
of each panel as the data to model ratio.

(a) A result with a CIE model and 3 gaussians.
(b) A fit with a CIE model, 3 gaussians, and a power law. The contribution of the CIE plasma and the power-law tail are plotted by red and green dashed lines, respectively. 
(c) The same as panel (b) but the power law is replaced by a cutoff power law (also plotted by green dashed line).
(d) A result with two CIE plasma model, plus 3 gaussians. The low and high temperature components are plotted by red and magenta dashed lines, respectively.
(e) A fit with the partially covered model (see \S\ref{subsection:thermal_interpretation}). The normally absorbed CIE model and power-law tail are plotted by red and green dashed lines, while highly absorbed components are indicated by magenta and cyanic dashed lines.
}
\label{fig:sifit_rega}
\end{figure*}

Figure \ref{fig:sifit_rega} shows results of the 
simultaneous fitting to the XIS and PIN spectra. 
Model (a) failed to reproduce the spectra with $\chi^2_\nu= 3.32~(740)$. 
The derived plasma temperature of $kT=9.2$ keV is much higher than the value of 
$6.4-6.6$ keV calculated based on the Fe line intensity ratio \citep{koy07a}. 
Moreover, the extrapolated thermal model falls 
significantly short of the observed PIN spectrum. 
All these results indicate that a separate hard tail component is necessary, 
thus reconfirming \citet{koy07a}.

Models (b) and (c) gave much more successful fits to the spectra, 
with $\chi^2_\nu$=1.52~(738) and $\chi^2_\nu$=1.01~(737), respectively.
As listed in table \ref{tab:sifitparameters},
the plasma temperatures derived with these models are 
both consistent with the above quoted XIS measurement.
However, Model (b) gives a photon index of $\Gamma=1.86$
which is steeper than the XIS determination,
and is over-predicting the counts in energies above $\sim25$ keV.
Model (c), with a photon index of $\Gamma=0.47$ 
and a cutoff energy of $E_\mathrm{c}=9.9$ keV,
better reproduce the XIS and PIN spectra. 
Therefore, the cutoff power-law modeling in Model (c) is considered 
more appropriate than the simple power law employed in Model (b),
in agreement with the result obtained using the PIN spectra alone
(\S~\ref{subsection:HXDfits}).

With Models (b) and (c),
the constant factor adjusting the model normalization to fit the PIN data
was obtained as 0.46 and 0.34, respectively. 
This means that the extended emission is brighter inside the XIS FOV, 
because a largely extended emission with uniform brightness would make this factor 1.0.

We repeated the same analysis by changing the iron abundance as 0.5 and 2.0, 
to find that the essential properties of the results 
obtained assuming the 1.0 abundance remain unchanged. At the same time,
we examined the uncertainty of 5\% in the PIN NXB modeling, and confirmed
the same results within the statistical errors.
In addition, the results were qualitatively unchanged 
when we varied the modeling of the bright (3 plus 11) point sources 
within plausible tolerance (e.g., changing $\Gamma_\mathrm{p}$ of ``the other'' sources or 
replacing a power law to a cutoff power law).
Therefore we conclude that the residual PIN spectra of Region A,
combined with that from the XIS, require
the hard tail other than a thermal component, 
and the hard tail exhibits a mildly convex shape in the higher energy band. 

\begin{table}
\caption{Results of simultaneous fits to the XIS and residual PIN spectra of Region A.\footnotemark[$*$]}\label{tab:sifitparameters}
\begin{center}
\begin{tabular}{lcc}
%%%%%
\hline\hline				
	&	(b)TH+PL\footnotemark[$\dagger$]	&	(c)TH+CUTPL\footnotemark[$\dagger$]	\\
	\hline
$N_\mathrm{H}$ ($\mathrm{cm}^{-2}$) 	&	$6\times10^{22}$ (fixed)	&	$6\times10^{22}$ (fixed)	\\
$N_\mathrm{Fe}$ ($\mathrm{cm}^{-2}$)	&	$6.2^{+0.6}_{-0.6}\times10^{18}$	&	$10.9^{+1.2}_{-0.3}\times10^{18}$	\vspace{2mm}\\
\hline					
\multicolumn{3}{l}{Therma}\\					
\hline
$kT$ (keV)	&	$7.0^{+0.1}_{-0.1}$	&	$6.2^{+0.1}_{-0.1}$	\\
$Z_\mathrm{Fe}$	(solar)&	1.0 (fixed)	&	1.0 (fixed)	\\
$Z_\mathrm{Ni}$	(solar)&	$2.1^{+0.3}_{-0.3}$ 	&	$2.2^{+0.3}_{-0.5}$	\\
Redshift	&	$1.8^{+0.1}_{-0.1}\times10^{-3}$ 	&	$1.9^{+1.1}_{-0.1}\times10^{-3}$ \\
Norm.\footnotemark[$\ddagger$]	&	$3.63^{+0.03}_{-0.03}$	&	$3.55^{+0.04}_{-0.04}$	\vspace{2mm}\\
\hline					
Emission lines & &\\					
\hline					
Fe \emissiontype{I}\footnotemark[$\S$] K$\alpha$ & & \\					
$E_{\rm center}$ (eV)	&	$6403^{+1}_{-1}$	&	$6401^{+1}_{-1}$	\\
$\sigma$ (eV)	&	$11^{+5}_{-11}$	&	$12^{+12}_{-8}$	\\
Intensity\footnotemark[$\|$]	&	$3.18^{+0.05}_{-0.05}\times10^{-2}$	&	$3.26^{+0.01}_{-0.01}\times10^{-2}$	\vspace{2mm}\\
Fe \emissiontype{I}\footnotemark[$\S$] K$\beta$	&		&		\\
$E_{\rm center}$\footnotemark[$\#$] (eV)	&	$7062$	&	$7060$	\\
$\sigma$\footnotemark[$\#$] (eV)	&	$12$	&	$13$	\\
Intensity\footnotemark[$\|$]	&	$4.2^{+0.1}_{-0.1}\times10^{-3}$	&	$5.6^{+0.3}_{-0.4}\times10^{-3}$	\vspace{2mm}\\
Ni \emissiontype{I}\footnotemark[$\S$] K$\alpha$	&		&		\\
$E_{\rm center}$ (eV)	&	$7486^{+21}_{-21}$	&	$7478^{+20}_{-16}$	\\
$\sigma$ (eV)	&	$0$ (fix)	&	$0$	(fix)\\
Intensity\footnotemark[$\|$]	&	$1.5^{+0.4}_{-0.4}\times10^{-3}$	&	$2.1^{+0.3}_{-0.6}\times10^{-3}$	\vspace{2mm}\\
\hline					
Hard tail & &\\					
\hline					
$\Gamma$	&	$1.86^{+0.01}_{-0.01}$	&	$0.47^{+0.01}_{-0.01}$	\\
$E_\mathrm{c}$ (keV)	&	---	&	$9.9^{+0.1}_{-0.1}$	\\
Norm.\footnotemark[$**$]	&	$8.21^{+0.15}_{-0.15}\times10^{-1}$	&	$1.66^{+0.02}_{-0.03}\times10^{-1}$	\\
\hline					
	&		&		\\
%Constant	&	$0.577^{+}_{-}$	&	$0.432^{+0.08}_{-0.05}$	\\
Constant	&	$0.45^{+0.01}_{-0.01}$	&	$0.34^{+0.06}_{-0.04}$	\\
& & \\
$\chi^2_\nu~(\nu)$	&	1.51 (738)	&	1.01 (737)	\\
\hline
%%%%%
\multicolumn{3}{@{}l@{}}{\hbox to 0pt{\parbox{85mm}{\footnotesize
\vspace{0.2cm}
\footnotemark[$*$]Errors are at 90\% confidence level.\par
\footnotemark[$\dagger$]TH+PL : a CIE plasma emission plus three gaussian lines with a power-law tail. TH+CUTPL : a CIE plasma emission plus three gaussian lines with a cutoff power-law tail.\par
\footnotemark[$\ddagger$]$10^{-14}/(4\pi D^2)\int n_\mathrm{e} n_\mathrm{H} dV$, where $D$ is the distance to the source (cm), $n_{\rm e}$ and $n_{\rm H}$ are the electron and hydrogen density in cm$^{-3}$, respectively.\par
\footnotemark[$\S$]Or in low ionization states.\par
\footnotemark[$\|$]In units of photons s$^{-1}$ cm$^{-2}$.\par
\footnotemark[$\#$]Fixed at 1.103 $\times$ $E_\mathrm{center}$(Fe \emissiontype{I} K$\alpha$) and 1.103 $\times$ $\sigma$(Fe \emissiontype{I} K$\alpha$).\par
\footnotemark[$**$]In units of photons s$^{-1}$ cm$^{-2}$ keV$^{-1}$ at 1 keV.\par
}\hss}}
\end{tabular}
\end{center}
\end{table}

\section{Discussion}

\subsection{Summary of the obtained results}
In the HXD-PIN  $10-40$ keV band,
we detected intense hard X-ray signals from all the five regions around the GC. 
Except for one case,
the PIN signal counts did not vary significantly within each observation (figure~\ref{fig:lc}).
When some regions were observed multiple times,
neither the hard X-ray intensity nor the spectral shape 
changed significantly (figure \ref{fig:gcsrc1}, figure~\ref{fig:gc5regions}).
The background-subtracted PIN spectra are approximated 
by a power-law model of $\Gamma= 2.0 -2.7$,
with a typical $12-40$ keV flux of $(3-4) \times 10^{-10}~\ergcms$
per PIN FOV (table \ref{tab:pin_powerlaw_hxdnom}).

In all cases, the background-subtracted PIN signals exceeded
what would be expected from the XIS data of the same pointings
(figure~\ref{fig:gcsrc1}, figure~\ref{fig:gc5regions}), 
due most likely to the difference between their fields of view.
Using the offset XIS observations (figure \ref{fig:offset}) and near-simulatateous INTEGRAL coverage,
we subtracted contributions of catalogued bright point sources 
(table \ref{tab:pslist}) that fall inside the PIN FOV.
The PIN signals were then typically halved (figure~\ref{fig:subtraction}),
but still remained significant up to $\sim 40$ keV 
(figure~\ref{fig:regionab_residual_spectra}).
Therefore, we consider that the hard X-ray emission is also as
extended as the thermal emission observed in $<10$ keV.
The derived residual PIN spectra  exhibit mildly convex shapes 
(figure \ref{fig:residual_spectra_of_regionab_fit}, 
 table \ref{tab:residual_spectra_of_regionab_fit}).

By fitting the XIS and PIN data of Region A simultaneously, 
we have shown that the broad-band (5-40~keV) spectrum of the extended X-ray emission
cannot be reproduced by a single CIE model plus gaussian lines, 
but requires an additional harder component 
which was first suggested by \citet{koy07a} (figure~\ref{fig:sifit_rega}).
This harder component is successfully reproduced by a
mildly curving cutoff power-law model,
while a single power law is less successful (table \ref{tab:sifitparameters}).
As judged from the derived PIN vs. XIS normalization,
the surface brightness of the hard emission is inferred to 
decrease toward the periphery of the PIN FOV,
rather than being uniform within it.

\subsection{Surface brightness distribution}\label{subsection:distribution}
To study a degree-scale surface brightness distribution
of the extended GC emission in the PIN band, 
we further  derived the residual PIN signals from Regions A$_3$, C, D, and E. 
In estimating and subtracting  contributions from  ``the other'' sources,
the method described in \S\ref{subsec:subtraction} 
was used assuming $\Gamma_\mathrm{p}=2.1$.
Although near simultaneous XIS data on the three point sources
are no longer available, we estimated their contributions
assuming that their spectral shape did not change from
the first offset XIS data acquired on 2005 September,
and setting their normalization parameters 
to reproduce the one-week-averaged INTEGRAL IBIS flux
obtained around the observation date of those regions.

\begin{figure}
  \begin{center}
    \FigureFile(80mm,50mm){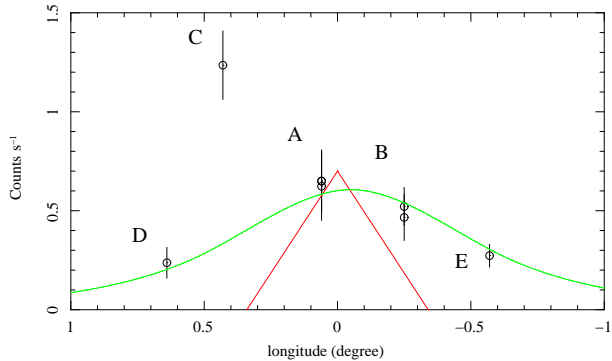}
  \end{center}
  \caption{The galactic longitude distributions of the apparently extended emission components. 
Black circles show the $12-40$ keV PIN count rates
obtained after subtracting the NXB and the contributions from bright point sources.
The error bars represent uncertainties of the 
estimation of fluxes of the bright point sources.
The green curve shows the 6.7 keV Fe line intensity distribution 
measured with ASCA GIS \citep{mae98}, 
convolved with  the PIN angular response
which is shown by a red triangle.
}
  \label{fig:countrate_l_distribution}
\end{figure}

The $12-40$ keV count rates of the residual PIN data, 
obtained in this way, are shown in figure \ref{fig:countrate_l_distribution} 
as a function of the galactic longitude. 
In all regions, the residual counts are thus positive. 
Furthermore, the emission detected with PIN
is spatially more extended than a point source,
as the measured longitude distribution is clearly wider
than the PIN angular response for a point source (a red triangle).

Using the ASCA GIS data, \citet{mae98} have shown 
that the surface brightness of 
the 6.7 keV Fe line around the GC decreases along the
longitudinal direction on two angular scales, $\timeform{0\circ.42}\pm\timeform{0\circ.06}$
and $\sim15^\circ$.
To compare the present hard X-ray results with the Fe line intensity distribution,
we modeled the latter as $\propto \exp[-(l+\timeform{0\circ.05})/\timeform{0\circ.42}]$,
and convolved it with the PIN angular response.
The larger angular scale of \citet{mae98} can be neglected, and
the offset of $l=\timeform{-0\circ.05}$ represents the peak structure of
the line emission around Sgr A$*$.

As presented in figure \ref{fig:countrate_l_distribution},
the convolved Fe line intensity distribution has come to a 
close agreement with the longitudinal PIN count-rate profile.
Even considering that the PIN signals are partially contributed
by the thermal component to which the Fe lines are associated,
the hard X-ray emission is inferred to have a similar
spatial distribution as the hot thermal emission.

\subsection{The case of Region C}\label{subsection:regionc}
As noted earlier,
Region C exhibited almost twice higher counts in the PIN band
than the other regions (figure \ref{fig:gc5regions}),
even though the XIS data of the region show no corresponding sign.
Furthermore, the PIN count rate varied significantly during this
particular pointing (figure \ref{fig:lc} bottom).

We consider that a transient source SAX J1747.0$-$2853, 
which was outside the XIS FOV but inside that of PIN, 
caused this discrepancy.
According to observations with Chandra, Swift, and INTEGRAL,
this transient was in a flaring state since 2005 October (e.g., \cite{wij05}; \cite{kon05}; \cite{kuu07}),
with an intensity of $\sim10$ mCrab in the $20-60$ keV band \citep{kuu07}.
It further brightened up to $\sim20$ mCrab during our Region C observation, 
according to the INTEGRAL IBIS monitoring data \citep{kuu07}.

In the estimation of the bright point source 
contributions (\S\ref{subsection:distribution}), 
we scaled the power-law model of SAX J1747.0$-$2853 to
the $20-60$ keV intensity of 20 mCrab reported by IBIS.
However, even after this subtraction, the Region C counts remained
unusually high (figure \ref{fig:countrate_l_distribution}). Supposing that the
proper Region C brightness is comparable to that of Region A,
the excess in figure \ref{fig:countrate_l_distribution} can be
explained if SAX J1747.0$-$2853 was during the Region C
observation twice as bright ($\sim40$ mCrab) as the week-averaged
IBIS intensity. Such a variation is reasonable for transient LMXBs.

%----------------------------5.4-------------------------------------------
\subsection{Comparison with the ridge and bulge emission}
%--------------------------------------------------------------------------
The extended X-ray emission, accompanied by a hard tail,
has been observed by many investigators from the Galactic ridge and bulge regions.
The spectrum is described  by one or two thermal component(s) plus a power-law tail; 
e.g. $kT=3.1\pm1.4$ keV and $\Gamma=1.58\pm1.05$ 
in the Galactic ridge region \citep{yam97}, 
while $kT\sim3$ keV and $\Gamma\sim1.8$  
in the Galactic bulge \citep{kok01}. 
In both cases, the thermal and power-law components 
cross are inferred over at $\sim6-8$ keV in the spectrum. 
It is remarkable that the two regions exhibit very similar spectra,
in spite of different interstellar environments.

The present Suzaku studies have for the first time clarified 
that such a hard-tail component also accompanies the extended  GC emission; 
it underlies the thermal emission in the XIS band,
and carries a dominant flux in the HXD-PIN band.
When this component is represented by a  cutoff power-law model,
it is inferred to  cross with  the thermal component
at an energy of $\sim 7$ keV (figure \ref{fig:sifit_rega}).
This crossing energy is very close to those found in the other two regions,
implying a close similarity among the extended X-ray emissions
from the three distinct spatial components. 

According to \citet{val00},
the $10-400$ keV hard X-ray spectrum of the Galactic  ridge emission 
obtained with the CGRO OSSE is represented better
by a cutoff power-law model  
($\Gamma=0.63\pm0.25,~ E_\mathrm{c}=41.4^{+13.0}_{-8.4}$~keV),
than by a single power law.
This appears similar to our cutoff power-law modeling of the GC hard X-rays.
However, some caution is needed in this analogy,
because the cutoff power-law  model reached by \citet{val00}
and that from the present work have quantitatively rather different slopes,
becoming discrepant by a factor of $\sim 8$ at 40 keV,
when they are equalized at 10 keV.
A more quantitative comparison must await 
observations of the GC emission over still wider  energies.

%--------------------------------- 5.5 ---------------------------------
\subsection{Thermal interpretation of the hard X-ray emisison}
\label{subsection:thermal_interpretation}
%---------------------------------- 5.5 -----------------------------------
After the work by \citet{val00}, 
\citet{kri07} argued that the hard X-ray ridge spectrum, with  a mildly convex shape,  
may be interpreted as high temperature thermal bremsstrahlung.
Then we also tried to reproduce the Region A spectra with a model consisting of two CIE plasma components and three gaussian lines (hereafter 2T model). 
The two CIE components were assumed to suffer a common 
absorption by  $6\times10^{22}~\mathrm{cm}^{-2}$,
and  were constrained to have the same (but free) metal abundances.
The model reproduced the spectrum well with $\chi^2_\nu=1.12~(736)$,
as presented in figure \ref{fig:sifit_rega}d.
The two CIE temperatures were obtained as $3.7^{+0.8}_{-0.2}$ keV and $17.8^{+1.0}_{-0.7}$ keV,
while the Fe and Ni abundances as $0.81^{+0.03}_{-0.02}$ and $2.0^{+0.3}_{-0.3}$ solar, respectively.
The best fit parameters are listed in table \ref{tab:sifitparameters_discussion}.
Although the ionized iron lines have been explained successfully by \citet{koy07a}
using a single CIE plasma with a temperature of $5-7$ keV,
the present 2T model explains them equally well 
as a superposition of the cooler and hotter CIE components,
which mainly account for the He-like and H-like lines, respectively.

The hotter CIE component  implied by the above 2T fit 
has such a high temperature (17.8 keV).
As argued by \citet{kri07},
the temperature would be interpreted easily
if the extended hard X-ray emission is composed of a numerous 
high-temperature thermal point-like sources, such as magnetic CVs.
In particular,  intermediate polars are good candidates,
because their  thermal emission sometimes becomes 
as hot as several tens of keV \citep{lam79}.
Furthermore, this interpretation is consistent with the 
existence of the strong fluorescent Fe K lines visible in 
our XIS spectra, because CVs (both magnetic and non-magnetic) are 
generally known to emit these lines as well 
(\cite{muk93}; \cite{ezu99}; \cite{ran06}).

Although the CV interpretation of the hard component
would  thus appear plausible,
it is at the same time subject to several problems.
For example, in order to explain  the very constant distribution 
of the 6.7 keV vs. 6.9 keV line  intensity ratio 
in the $\timeform{-0\circ.4}<l<\timeform{0\circ.6}$ region (\cite{koy07a}; \cite{nob07}),
the assumed hard point sources must have nearly 
the same spatial distribution as the cooler CIE emitter.
Yet another difficulty  is
that the 6.7 keV Fe line intensity of the GC emission as measured with the XIS 
distributes asymmetrically between $l>0$ and $l<0$ \citep{koy07a},
which is absent in that of X-ray point sources down to a $2-8$ keV flux of $3\times10^{-15}\ergcms$ \citep{mun03}.
Furthermore, according to the 2T fit,
the 6.4 keV Fe K line is inferred to have an equivalent width (EW) of
$\sim600$~eV against the hotter CIE component,
while those of  CVs are much smaller ($\sim 50-200$ eV; \cite{ezu99}; \cite{ran06}).
Therefore, we must invoke a separate, or an additional,
source of the fluorescent lines from the GC region.

\begin{table}
\caption{Results of simultaneous fits to the XIS and residual PIN spectra of Region A with the 2T model and the partially covering model.\footnotemark[$*$]}\label{tab:sifitparameters_discussion}
\begin{center}
\begin{tabular}{lccc}
%%%%%
\hline\hline					
	&	(d)~2T\footnotemark[$\dagger$]	&	(e)~P.C.\footnotemark[$\dagger$]		\\
\hline
$N_\mathrm{H}$ (cm$^{-2}$)	&	$6\times10^{22}$~(fix)	&	$6\times10^{22}$~(fix)	\\
$N^*_\mathrm{H}$ (cm$^{-2}$)	&	$-$	&	$30\times10^{22}$~(fix)	\\
$N_\mathrm{Fe}$ (cm$^{-2}$)	&	$8.8^{+0.9}_{-1.0}\times10^{18}$	&	$8.0^{+0.8}_{-0.8}\times10^{18}$	\vspace{2mm}	\\
\hline					
Thermal	&				\\
\hline					
$kT_1$ (keV)	&	$3.7^{+0.8}_{-0.2}$	&	$6.1^{+0.2}_{-0.1}$	\\
$Z_\mathrm{Fe}$ (solar)	&	$0.81^{+0.03}_{-0.02}$	&	1.0~(fixed)	\\
$Z_\mathrm{Ni}$ (solar)	&	$2.0^{+0.3}_{-0.3}$	&	$2.1^{+0.4}_{-0.4}$	\\
Redshift	&	$1.31^{+0.03}_{-0.15}\times10^{-3}$	&	$1.9^{+0.1}_{-0.1}\times10^{-3}$	\\
Norm.$_1$\footnotemark[$\ddagger$]	&	$4.33^{+0.16}_{-0.12}$	&	$2.56^{+0.13}_{-0.04}$	\\
$kT_2$ (keV)	&	$17.8^{+1.0}_{-0.7}$	&	$-$	\\
Norm.$_2$\footnotemark[$\ddagger$]	&	$3.36^{+0.12}_{-0.21}$	&	$-$	\vspace{2mm}	\\
\hline					
Emission lines	&				\\
\hline					
Fe \emissiontype{I}\footnotemark[$\S$] K$\alpha$ &\\
$E_{\rm center}$ (eV)	&	$6402^{+1}_{-1}$	&	$6400^{+2}_{-2}$	\\
$\sigma$ (eV)	&	$<11$	&	$<6$	\\
Intensity\footnotemark[$\|$]	&	$3.50^{+0.05}_{-0.10}\times10^{-2}$	&	$4.94^{+0.06}_{-0.05}\times10^{-2}$	\\
Fe \emissiontype{I}\footnotemark[$\S$] K$\beta$					\\
$E_{\rm center}$\footnotemark[$\#$] (eV)	&	$7061$	&	$7059$	\\
$\sigma$\footnotemark[$\#$] (eV)	&	$<12$	&	$<7$	\\
Intensity\footnotemark[$\|$]	&	$4.7^{+0.4}_{-0.4}\times10^{-3}$	&	$3.6^{+0.3}_{-0.3}\times10^{-3}$	\\
Ni \emissiontype{I}\footnotemark[$\S$] K$\alpha$					\\
$E_{\rm center}$ (eV)	&	$7482^{+20}_{-19}$	&	$7479^{+19}_{-19}$	\\
$\sigma$ (eV)	&	$0$ (fix)	&	$0$ (fix)	\\
Intensity\footnotemark[$\|$]	&	$1.8^{+0.4}_{-0.4}\times10^{-3}$	&	$1.4^{+0.4}_{-0.3}\times10^{-3}$\vspace{2mm}	\\
\hline					
Hardtail	&				\\
\hline					
$\Gamma$	&	$-$	&	$2.31^{+0.03}_{-0.03}$	\\
Norm.\footnotemark[$**$]	&	$-$	&	$2.39^{+0.06}_{-0.08}$\vspace{2mm}		\\
\hline					
%Constant	&	$0.49786^{+0.0106}_{-0.00702}$	&	$0.49618^{+0.01649}_{-0.01708}$	\\
Constant	&	$0.39^{+0.01}_{-0.01}$	&	$0.39^{+0.01}_{-0.01}$	\\
$f$\footnotemark[$\dagger\dagger$]	&	$-$	&	$0.56^{+0.05}_{-0.13}$	\\
	&		&		\\
$\chi^2_\nu~(\nu)$	&	1.12~(736)	&	1.14~(736)	\\
\hline
\multicolumn{3}{@{}l@{}}{\hbox to 0pt{\parbox{85mm}{\footnotesize
\vspace{0.2cm}
\footnotemark[$*$]Errors are at 90\% confidence level.\par
\footnotemark[$\dagger$]2T : two thermal plasma emission plus three gaussian lines. P.C. : a thermal emission model with a power-law tail partially covered with a dense absorbing matter.\par
\footnotemark[$\ddagger$]$10^{-14}/(4\pi D^2)\int n_\mathrm{e} n_\mathrm{H} dV$, where $D$ is the distance to the source (cm), $n_{\rm e}$ and $n_{\rm H}$ are the electron and hydrogen density in cm$^{-3}$, respectively.\par
\footnotemark[$\S$]Or in low ionization states.\par
\footnotemark[$\|$]In units of photons s$^{-1}$ cm$^{-2}$.\par
\footnotemark[$\#$]Fixed at 1.103 $\times$ $E_\mathrm{center}$(Fe \emissiontype{I} K$\alpha$) and 1.103 $\times$ $\sigma$(Fe \emissiontype{I} K$\alpha$).\par
\footnotemark[$**$]In units of photons s$^{-1}$ cm$^{-2}$ keV$^{-1}$ at 1 keV.\par
\footnotemark[$\dagger\dagger$]Fraction of non-covered component in the P.C. model.\par
}\hss}}				
%%%%%
\end{tabular}
\end{center}
\end{table}

%---------------------------------- 5.6 -----------------------------------
\subsection{Non-thermal interpretation of 
the hard X-ray emission with a partially covered model}
%---------------------------------- 5.6 -----------------------------------
The thermal plus cutoff power-law model, fitted simultaneously to the XIS and PIN spectra, yielded a photon index of 0.47 (table \ref{tab:sifitparameters}) and
a cutoff energy of 9.9 keV. This small photon index suggests the presence of a highly absorbed component in the spectrum.
In addition, the GC region hosts a number of dense molecular clouds,
with the line-of-sight absorbing  column often exceeding $10^{23}~\mathrm{cm}^{-2}$.
Therefore, it is plausible to assume that
the strong absorption modifies the incoming continuum of the extended GC emission  
(either diffuse or point-source assembly), by reprocessing and partially absorbing it.
These processes are expected to make the observed spectrum slightly convex,
because the continuum in the XIS range will flatten.
This idea is consistent with the fact
that the PIN spectrum from Region A, where clouds must be more plenty, demands a convex model 
more strongly than that from Region B (table \ref{tab:residual_spectra_of_regionab_fit}).
Such dense clouds will not only change the continuum shape,
but will also produce the intense fluorescent lines
when they are hit by the hard GC X-rays 
(regardless of its origin),
and/or by other excitation sources  
such as relativistic particles \citep{byk02},
or past activity of the GC black hole \citep{koy96}.
Actually, dense molecular clouds are known to
emit intense 6.4 keV Fe K$\alpha$ line \citep{koy96}.

In order to examine
whether the mild spectral cutoff seen in the PIN energy band 
can be explained by the presence of dense interstellar molecular clouds,
we  introduced an alternative model to fit  the combined spectra of Region A.
It is the same as Model (b) of section~\ref{subsection:sifit}
(consisting of a CIE  model,  three gaussians, and a power-law tail),
but all the model components are now subjected to 
a partially-covered  absorption (using \texttt{wabs}).
That is, a certain fraction $f$ (left free to vary) of the overall spectral model
is assumed to be absorbed only by 
$N_\mathrm{H}=6\times10^{22}~\mathrm{cm}^{-2}$,
while the rest, $(1-f)$, is covered additionally by a thicker column of $N_{\rm H}^*$.
For the same reason as  in \S\ref{subsection:sifit}, 
we fixed the Fe abundance of the CIE plasma model to 1.0 solar,
and furthermore, 
$N_{\rm H}^*$ to a representative value of $30\times10^{22}~\mathrm{cm}^{-2}$
because it is not well constrained.
As shown in figure \ref{fig:sifit_rega}e,
this model has indeed given an equally acceptable fit 
as the 2T model, with $\chi^2_\nu=1.14~(736)$.
The determined plasma temperature and photon index are 
$6.1^{+0.2}_{-0.1}$ keV and $2.31^{+0.03}_{-0.03}$, 
while the non-covered fraction is  $f=0.56^{+0.05}_{-0.13}$.
The best fit parameters are listed in table \ref{tab:sifitparameters_discussion}.
We repeated the fitting by  changing   $N_{\rm H}^*$
over $(10-50)\times10^{22}~\mathrm{cm}^{-2}$, 
and obtained qualitatively the same results.

The EW of the 6.4 keV Fe line emission turned out to be $\sim540$ and $\sim270$~eV, against the strongly absorbed continuum and the total (strongly absorbed plus mildly absorbed) continuum, respectively.
According to \citet{mak86}, reprocessing of a power-law ($\Gamma=0.8$) emission by a spherically surrounding matter, with a column density of $N_\mathrm{H}=30\times10^{22}$~cm$^{-2}$ and cosmic abundance, is expected to produce a fluorescent Fe line with EW$\sim300$~eV (Model II of \cite{mak86}). If the surrounding matter is somehow transparent only toward our line of sight (Model I of the same article), the EW decreases to 200 eV.
In the present case, the direct and strongly absorbed components are mixed with the fraction $f$, so that the derived EW ($270-540$ eV) should be compared with the mixture of the above two cases (300 and 200 eV).
Considering that the extended GC emission has a spectral shape different from the power law ($\Gamma=0.8$) assumed in \citet{mak86},
and that the molecular clouds in the GC region may be richer in heavy elements, 
the agreement (within a factor of $\sim2$) is considered tolerable.

Introducing the above partial-covering model, 
we have successfully reconciled the mildly curving PIN spectra
with the results from the detailed plasma diagnostics,
obtained by \citet{koy07a} using the same XIS data.
Specifically, the CIE plasma model can be attributed to thermal emission 
from a truly diffuse hot plasma filling the GC region;
in particular, the obtained CIE temperature agree with 
that derived by \citet{koy07a}.
The power-law hard tail with $\Gamma=2.3$
can be comfortably interpreted as  diffuse non-thermal emission 
from accelerated non-thermal or supra-thermal electrons,
as argued for the ridge emission by, e.g., 
\citet{yam97}, \citet{dog02}, and \citet{mas02},
and for the bulge emission by \citet{kok01}.
The value of $f$ implies 
that the  covered and uncovered fractions are roughly comparable,
in agreement with our picture
invoking the diffuse hard-tail emitter intermixed with molecular clouds.

%\section{Summary}
%The five regions around the GC have been observed with Suzaku. We detected hard X-ray emission up to 40 keV with HXD-PIN and the emission is apparently extended over the FOV of the HXD and XIS. We estimated and subtracted the contribution of the catalogued bright ($\gtrsim1$~mCrab) point sources to the PIN signals and about halves of the original PIN signals remained after the subtraction. 
%Fitting the combined XIS and residual PIN spectra, we confirmed that the GC extended emission consists of a thermal emission accompanied with a power-law-like hard tail. The hard tail component favored a slightly convex shape rather than a simple power law. The wide-band spectrum can also be explained by the two thermal emission model or the thermal plus power law with strong 

%\begin{itemize}
%\item GCの周り5領域を観測
%\item 5領域全てからXISバンドだけでなくPINバンドでも強いemissionを検出
%\item そのスペクトルは、(明るい点源の寄与はあるが)広がった放射の存在を示唆
%\item 同時フィットからは、thermal成分だけでは足りないことがわかり、Koyama et al. 2007の結果をconfirm。高エネルギー側でとくに効いてくるhard tail成分は、今回の点源見積もりを用いると、若干折れ曲がっている
%\item 折れ曲がったスペクトルはIPに典型的な、〜20 keVという高温のthermal放射でもフィットできるが、部分的に強い吸収を受けた(thermal+pow)モデルでもフィットできる
%\end{itemize}

%謝辞

\end{document}